\documentclass[aps,pra,reprint,superscriptaddress,floatfix]{revtex4-2}

\usepackage{booktabs} 
\usepackage{amsmath,amssymb,amsfonts}
\usepackage{algorithmic}
\usepackage{graphicx}
\usepackage{xcolor}
\usepackage{listings}
\usepackage{float}
\usepackage{tikz-timing}
\usepackage{braket}
\usepackage{siunitx}
\usepackage{accents}
\usepackage{needspace}
\usepackage{placeins}
\usepackage{comment}
\usepackage[colorlinks=true,allcolors=blue]{hyperref}
\usepackage{setspace}
\usepackage{tikz}
\usetikzlibrary{arrows.meta, positioning, fit, shapes.geometric}

\usepackage{dcolumn}
\usepackage{bm}

\lstdefinestyle{spectre}{
  basicstyle=\ttfamily\small,
  columns=fullflexible,
  breaklines=true,
  breakatwhitespace=true,
  frame=none,
  framerule=0.4pt,
  framesep=3pt,
  xleftmargin=0pt,
  xrightmargin=3pt,
  resetmargins=true,
  keepspaces=false,
  showstringspaces=false,
  aboveskip=0.5\baselineskip,
  belowskip=0.5\baselineskip
}

\begin{document}
\preprint{APS}

\title{Valley-Aware Optimal Control of Spin Shuttling Using Cryogenic Integrated Electronics}

\affiliation{Integrated Computing Architectures (PGI-4), Forschungszentrum Jülich, 52425 Jülich, Germany}
\affiliation{Quantum Control (PGI-8), Forschungszentrum Jülich, 52425 Jülich, Germany}
\affiliation{Institute for Theoretical Physics, University of Cologne, 50937 Cologne, Germany}
\affiliation{Faculty of Engineering, Communication Systems, University of Duisburg-Essen, 47057 Duisburg, Germany}
\affiliation{System Engineering for Quantum Computing, RWTH Aachen University, 52056 Aachen, Germany}
\affiliation{IceCirc GmbH, 41844, Wegberg, Germany}

\author{Pau Dietz Romero}
\email{p.dietz.romero@fz-juelich.de}
\affiliation{Integrated Computing Architectures (PGI-4), Forschungszentrum Jülich, 52425 Jülich, Germany}
\affiliation{IceCirc GmbH, 41844, Wegberg, Germany}

\author{Nermine Chaabani}
\email{n.chaabani@fz-juelich.de}
\thanks{The first two authors contributed equally to this work.}
\affiliation{Quantum Control (PGI-8), Forschungszentrum Jülich, 52425 Jülich, Germany}
\affiliation{Institute for Theoretical Physics, University of Cologne, 50937 Cologne, Germany}

\author{Lammert Duipmans}
\affiliation{Integrated Computing Architectures (PGI-4), Forschungszentrum Jülich, 52425 Jülich, Germany}

\author{Alessandro David}
\affiliation{Quantum Control (PGI-8), Forschungszentrum Jülich, 52425 Jülich, Germany}

\author{Felix Motzoi}
\affiliation{Quantum Control (PGI-8), Forschungszentrum Jülich, 52425 Jülich, Germany}
\affiliation{Institute for Theoretical Physics, University of Cologne, 50937 Cologne, Germany}

\author{Stefan van Waasen}
\affiliation{Integrated Computing Architectures (PGI-4), Forschungszentrum Jülich, 52425 Jülich, Germany}
\affiliation{Faculty of Engineering, Communication Systems, University of Duisburg-Essen, 47057 Duisburg, Germany}
\affiliation{IceCirc GmbH, 41844, Wegberg, Germany}

\author{Lotte Geck}
\affiliation{Integrated Computing Architectures (PGI-4), Forschungszentrum Jülich, 52425 Jülich, Germany}
\affiliation{System Engineering for Quantum Computing, RWTH Aachen University, 52056 Aachen, Germany}
\affiliation{IceCirc GmbH, 41844, Wegberg, Germany}

\begin{abstract}
Electron shuttling is emerging as a key mechanism for enabling long-range coupling in scalable spin-qubit architectures. Bringing shuttling waveform generation into the cryostat can improve scalability, but imposes strict area and power constraints on the control electronics. Concurrently, shuttling in Si/SiGe is further limited by a spatially varying valley splitting that induces spin--valley mixing and degrades coherence. Here, we make three contributions that address these limitations jointly: (i) an end-to-end co-simulation framework that combines disorder-informed valley maps with transistor-level cryogenic circuit simulations including electronic noise; (ii) a fully integrated cryogenic shuttling-signal generator tailored to velocity modulation, enabling period-wise waveform shaping through discrete circuit settings stored in on-chip memory; and (iii) a noise-aware optimization procedure that tunes only these implementable circuit controls, using one of four discrete resistor settings per period, to generate high-fidelity shuttling sequences. 
Across simulated valley and noise realizations in our co-simulation framework, the optimized velocity-modulation waveforms improve transport performance, achieving an average shuttling fidelity of $99.99 \pm 0.007\%$ at $v_{\mathrm{avg}} = 20~\mathrm{m\,s^{-1}}$ over a distance of $10~\mu\mathrm{m}$, while maintaining active analog power consumption in the tens of $\mu\mathrm{W}$ during shuttling. This validates on-chip storage and replay of optimized control settings as a practical strategy to mitigate valley disorder in scalable shuttling architectures.
\end{abstract}

\maketitle

\section{Introduction}
\label{sec:introduction}

\subsection{Scalable spin qubit architectures and the role of shuttling}
\label{subsec:scalable_architecture}
Semiconducting spin qubits offer a route to scalable, fault-tolerant quantum computing by combining electrical controllability and readout \cite{kawakamiElectricalControlLonglived2014} with mature complementary metal–oxide–semiconductor (CMOS) fabrication techniques that enable integration with control electronics \cite{xueCMOSbasedCryogenicControl2021, vandersypenInterfacingSpinQubits2017, geckControlElectronicsSemiconductor2019, volmerMappingValleySplitting2024}.

Proposed scalable architectures, such as the SpinBus architecture \cite{kunneSpinBusArchitectureScaling2024} and the spiderweb array \cite{boterSpiderwebArraySparse2022}, pursue modular designs in which qubit units are arranged in sparse planar grids to enable scalable two-dimensional layouts. However, placing qubits in such sparse layouts requires coherent coupling over micrometer-scale distances between qubit sites. Due to the no-cloning theorem \cite{woottersSingleQuantumCannot1982}, quantum information must be transferred either through sequences of swap gates or by physically transporting the qubit. The latter, commonly referred to as shuttling, is employed in semiconductor spin-qubit~\cite{desmetHighfidelitySinglespinShuttling2025,struckSpinEPRpairSeparationConveyormode2024, xueSiSiGeQuBus2024,davidLongDistanceSpin2024, vanriggelen-doelmanCoherentSpinQubit2024}, neutral-atom~\cite{cicaliFastNeutralatomTransport2025, hwangFastReliableAtom2025, hauckEnhancedShortcutsAdiabaticity2026} and trapped-ion~\cite{anShortcutsAdiabaticityCounterdiabatic2016, hensingerTjunctionIonTrap2006, kaushalShuttlingbasedTrappedionQuantum2020} platforms.
In conveyor-mode shuttling \cite{desmetHighfidelitySinglespinShuttling2025, seidlerConveyormodeSingleelectronShuttling2022, xueSiSiGeQuBus2024, ermoneitOptimalControlConveyorMode2023}, the electron spin qubit is confined in a gate-defined electrostatic potential minimum within a Si/SiGe heterostructure. Periodic, phase-shifted gate voltages generate this traveling electrostatic potential, thereby confining and transporting the qubit over the shuttling channel \cite{langrockBlueprintScalableSpin2023, vandersypenInterfacingSpinQubits2017}. Furthermore, high-fidelity single-qubit gates can be realized with optimized shuttling control in dedicated manipulation zones equipped with a micro-magnet, where the electron is oscillated to drive electric-dipole spin resonance (EDSR) \cite{pazhedathLargeSpinshuttlingOscillations2025}.

In experimental demonstrations, conveyor-mode shuttling has been benchmarked at high performance using waveform generation at room temperature. Xue \textit{et al.}~\cite{xueSiSiGeQuBus2024} report a charge shuttling fidelity of $99.7 \pm 0.3\,\%$ in a Si/SiGe conveyor device. De Smet \textit{et al.}~\cite{desmetHighfidelitySinglespinShuttling2025} demonstrated spin-coherent shuttling with a fidelity of $99.54 \pm 0.05\,\%$. Struck \textit{et al.}~\cite{struckSpinEPRpairSeparationConveyormode2024} investigated spin coherence during conveyor-mode shuttling in a Si/SiGe heterostructure by separating and recombining an entangled electron pair prepared in a singlet--triplet (ST$_0$) state.
While these results validate conveyor-mode shuttling experimentally, scalable architectures motivate shifting waveform generation to integrated cryogenic electronics.

\subsection{Cryogenic control electronics for scalable shuttling}
\label{subsec:scalable_electronics}

Semiconductor heterostructures for spin qubits can be manufactured with commercial CMOS processes \cite{zwerverQubitsMadeAdvanced2022}, opening the perspective of lithographically scaling qubits together with integrated circuitry.
Integrating control electronics in close proximity to semiconductor spin qubits at the millikelvin stage reduces the number of interconnects and feedthroughs entering the cryostat, replacing them with microfabricated interconnects \cite{geckControlElectronicsSemiconductor2019}. 
Due to their smaller footprint compared to bond pads, such interconnects reduce wiring density and help alleviate the fan-out problem. 
Moreover, reducing the number of cables limits passive thermal links to higher temperature stages and avoids power dissipation in attenuators that would otherwise heat the coldest stages of the dilution refrigerator \cite{vandersypenInterfacingSpinQubits2017}.

In trapped ion quantum computing, the integration of digital-to-analog converters (DACs) for shuttling operations has been demonstrated \cite{meyerCryogenicEvaluationDigitaltoAnalog2025, siebererCryogenicDigitaltoAnalogConverter2024, stuartChipIntegratedVoltageSources2019}, typically relying on high-resolution, high-bandwidth waveform generation with power dissipation levels that are incompatible with operation at the millikelvin stage of dilution refrigerators.
Thus integrated electronics for spin qubits operating at mK require tailored, co-designed circuitry that minimizes both circuit area and power consumption \cite{geckControlElectronicsSemiconductor2019}.

These constraints motivate alternative approaches to shuttling-signal generation that avoid high-resolution, high-bandwidth waveform synthesis. 
One approach to improving the power and area efficiency of shuttling-signal generation is to relax the requirement for ideal sinusoidal waveforms, which are typically generated by room-temperature arbitrary waveform generators (AWGs) in current experiments \cite{seidlerConveyormodeSingleelectronShuttling2022, desmetHighfidelitySinglespinShuttling2025}. 
Instead, the cryogenic generation of periodic, smooth but non-sinusoidal waveforms has been proposed and simulated, achieving power consumption on the order of a few microwatts per signal \cite{nagaiDigitallyControlledConveyorbelt2025, duipmansPulseGenerationSpinQubit}, compatible with millikelvin operation.

\subsection{Valley-aware shuttling with cryogenic electronics}
\label{subsec:valley_aware_shuttling}

Nagai et al. \cite{nagaiDigitallyControlledConveyorbelt2025} proposed a cryogenic shuttling signal generator based on a discrete switch matrix and low-pass filtering, enabling signal generation at millikelvin temperatures. Their analysis assumes a uniform and sufficiently large valley splitting, whereas realistic Si/SiGe heterostructures often exhibit significant spatial variations in valley splitting due to interface disorder \cite{paqueletwuetzAtomicFluctuationsLifting2022, losertStrategiesEnhancingSpinShuttling2024, davidLongDistanceSpin2024}.

One of the central challenges for electron shuttling in Si/SiGe heterostructures is the strong spatial variation of the valley splitting along the shuttling path, i.e., the energy separation between the two lowest conduction-band valleys. Random alloy disorder at the Si/SiGe interface leads to a broad distribution of valley splittings, including a significant fraction of regions with valley splittings below $20~\mu\mathrm{eV}$. Because the two valley eigenstates are associated with different effective $g$ factors and therefore different spin-precession frequencies, this coupling leads to spin--valley entanglement during transport, resulting in dephasing and loss of spin purity. A range of practical device-level strategies has been proposed to enhance the valley-splitting magnitude~\cite{losertPracticalStrategiesEnhancing2023, thayilOptimizationSiSiGe2025}. Yet spatial fluctuations along the shuttling path can remain, so complementary control strategies are needed during transport. Several control strategies to mitigate valley-induced errors during shuttling have been proposed~\cite{losertStrategiesEnhancingSpinShuttling2024}, including velocity-modulation approaches~\cite{davidLongDistanceSpin2024}. However, the practical integration of such strategies with realistic cryogenic control electronics remains largely unexplored.

While the work of Nagai et al.~\cite{nagaiDigitallyControlledConveyorbelt2025} proposed and analyzed cryogenic shuttling signal generation using discrete component approaches under idealized assumptions on the valley landscape, the combined impact of realistic valley disorder and electronic non-idealities has not yet been investigated. In particular, the interplay between spatially varying valley splitting and circuit-level noise is expected to play a critical role in determining shuttling fidelity in scalable architectures based on integrated cryogenic electronics.

In this work, we close this gap with a hardware-embedded, valley-aware optimal-control approach for spin shuttling in which the control variables are defined by an integrated cryogenic circuit. We develop an end-to-end co-simulation framework that combines disorder-informed valley maps with transistor-level cryogenic circuit simulations, including electronic noise, and optimize shuttling waveforms under hardware-realizable constraints. A central contribution is an integrated cryogenic signal generator for shuttling, tailored to enable velocity modulation: it generates the phase-shifted traveling-wave shuttling drive and enables period-wise ramp shaping via compact, discrete circuit settings, avoiding high-resolution DAC synthesis and continuous waveform streaming into the cryostat. To the best of our knowledge, such a joint treatment of valley disorder, circuit non-idealities, and hardware-embedded optimal control within an integrated cryogenic shuttling signal chain has not been reported previously.

\section{Co-Simulation Methodology}
\label{sec:methodology}

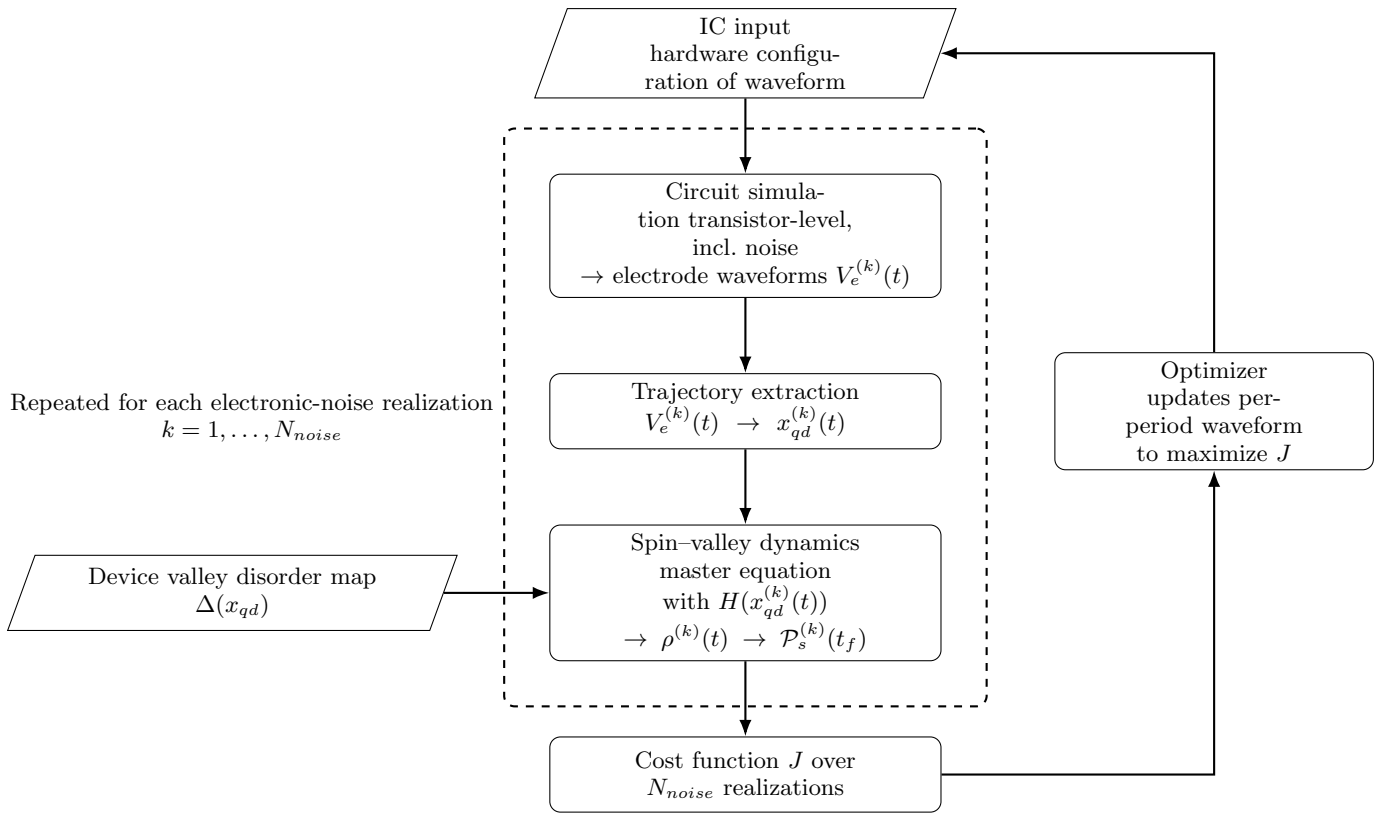
\begin{figure*}[t]
\centering
\begin{tikzpicture}[
  font=\small,
  node distance=10mm and 14mm,
  box/.style={draw, rounded corners, align=center, minimum width=34mm, minimum height=10mm,
  , text width=40mm},
  widebox/.style={draw, rounded corners, align=center, minimum width=52mm, minimum height=10mm, text width=45mm},
  io/.style={draw, trapezium, trapezium left angle=70, trapezium right angle=110, align=center, minimum width=34mm, minimum height=10mm, text width=45mm},
  arrow/.style={-{Latex[length=2.2mm]}, thick},
]

% Control/config
\node[io] (config) {IC input\\ hardware configuration of waveform};

% Circuit simulation
\node[widebox, below=of config] (spectre) {Circuit simulation transistor-level,\\
incl.\ noise\\
$\rightarrow$ electrode waveforms $V_e^{(k)}(t)$};

% Position extraction
\node[widebox, below=of spectre] (mapping) {Trajectory extraction\\
$V_e^{(k)}(t)\rightarrow x_{qd}^{(k)}(t)$};

% Quantum dynamics
\node[widebox, below=of mapping] (quantum) {Spin--valley dynamics\\
master equation with $H(x_{qd}^{(k)}(t))$\\
$\rightarrow \rho^{(k)}(t)\rightarrow \mathcal{P}_s^{(k)}(t_f)$};

\node[io, left=of quantum] (params) {Device valley disorder map\\ $\Delta(x_{qd})$};

% Metrics
\node[widebox, below=of quantum] (metrics) {Cost function $J$ over $N_{noise}$ realizations};

% Optimizer
\node[box, right=15mm of mapping] (opt) {Optimizer\\ updates per-period waveform\\ to maximize $J$};

\node[draw, dashed, rounded corners, thick, inner sep=6mm, fit= (spectre) (mapping) (quantum),
  label={[align=center]left: Repeated for each electronic-noise realization\\ $k = 1,\dots,N_{noise}$} ] (ensemblebox) {};

% Arrows main flow
\draw[arrow] (params) -- (quantum);
\draw[arrow] (config) -- (spectre);
\draw[arrow] (spectre) -- (mapping);
\draw[arrow] (mapping) -- (quantum);
\draw[arrow] (quantum) -- (metrics);

% Optimization loop
\draw[arrow] (metrics) -| (opt);
\draw[arrow] (opt) |- (config);

\end{tikzpicture}
\caption{End-to-end co-simulation workflow used to optimize and validate shuttling operations. The cryogenic signal generator is simulated at transistor level (\emph{Cadence Spectre}) to obtain gate voltages $V_e^{(k)}(t)$ for electronic-noise realization $k$. The waveforms are mapped to a shuttling trajectory $x_{qd}^{(k)}(t)$ and used to simulate spin--valley dynamics, yielding the final spin purity $P_s^{(k)}(t_f)$. An optimizer updates period-wise control parameters (one discrete configuration per signal period) to maximize $J$.}

\label{fig:cosim_workflow}
\end{figure*}

This work employs an end-to-end co-simulation framework to study and optimize spin qubit shuttling in Si/SiGe heterostructures under realistic cryogenic-electronics constraints. Its purpose is to enable a system-level assessment of shuttling fidelity that simultaneously accounts for valley disorder, electronic non-idealities, and hardware-embedded control.

The workflow comprises four stages: (i) transistor-level simulation of the cryogenic signal generator, (ii) extraction of the effective shuttling trajectory from the gate-voltage waveforms, (iii) evaluation of the resulting spin--valley dynamics during transport, and (iv) optimization of the hardware control parameters. A schematic overview is shown in Fig.~\ref{fig:cosim_workflow}.

In the first stage, the cryogenic shuttling signal generator described in Subsec.~\ref{subsec:circuit} is simulated at the transistor level, with electronic noise optionally enabled. This produces time-dependent voltage waveforms $V_e^{(k)}(t)$ applied to the gate electrodes. The co-simulation is performed in batches of $N_{noise}$ simulations that employ different noise realizations $k=1,\dots,N_{noise}$ to capture non-deterministic waveform variations due to electronic noise.

In the second stage, the simulated gate voltages are mapped to a noise-dependent trajectory $x_{qd}^{(k)}(t)$ of the moving quantum dot along the shuttling channel as described in Subsec.~\ref{subsec:qubit_position}. This trajectory determines how the electron samples the spatially varying valley landscape during transport and therefore defines the time dependence of the spin--valley Hamiltonian.

In the third stage, the extracted trajectory determines how the disorder-informed valley map $\Delta(x_{qd})$ is sampled, which in turn defines the spin--valley dynamics during shuttling. The resulting time-dependent Hamiltonian governs the evolution of the joint spin--valley system, which is simulated using an open-system description. From the resulting density matrix, figures of merit such as the final spin purity and the shuttling fidelity are obtained, allowing the quantification of residual spin--valley entanglement and decoherence induced during transport. The valley-disorder model, Hamiltonian, and open-system dynamics are described in detail in Subsecs.~\ref{subsec:valley_disorder}--\ref{subsec:open_system}.

In the final stage, the control parameters of the cryogenic signal generator are optimized using the method described in Sec.~\ref{sec:optimizer}. The optimization is generally performed in the presence of electronic noise to reflect the experimentally relevant environment, although the noise can be disabled for controlled analysis or exploration of deterministic effects. The resulting control sequences are then evaluated over multiple noise realizations to quantify their robustness against non-deterministic waveform fluctuations. Importantly, the optimization acts directly on hardware-compatible control parameters, enabling optimal control at the level of the integrated cryogenic electronics rather than through idealized waveforms.

The end-to-end co-simulation workflow presented here builds on the framework introduced by Dietz Romero \emph{et al.}~\cite{dietzromeroCoSimulationAutomatedOptimization2025}. In contrast to approaches that embed quantum-dynamics models directly into circuit simulators using equivalent circuits~\cite{borgarinoCircuitBasedCompactModel2022, periCompactQuantumDot2025, gysCircuitModelEfficient2021}, our implementation couples \emph{Cadence Spectre} to a Python-based qubit model via the Python interface \texttt{pyspectre}~\cite{uhlmannPySpectre2026}. This separation avoids the overhead of translating quantum dynamics into electronic circuit representations and facilitates interdisciplinary collaboration by preserving native abstractions for both electronic engineers and quantum physicists.

\section{Physical Model}
\label{sec:physical_model}
\subsection{Valley disorder along the shuttling trajectory}
\label{subsec:valley_disorder}
 A single electron is confined within a quantum dot and moved along the shuttling trajectory $x_{qd}$, thus sampling various regions of the Si/SiGe heterostructure interface. The atomic-scale variation of the interface becomes an intrinsic source of static disorder, leading to a position-dependent intervalley coupling $\Delta(x_{qd})$ defined as

\begin{equation}
\Delta(x_{qd})=\Delta_r(x_{qd})+i\,\Delta_i(x_{qd}),
\end{equation}
The local valley splitting, which represents the local energy separation between the two valley eigenstates, is defined as
\begin{equation}
E_v(x_{qd}) \equiv 2\lvert \Delta(x_{qd}) \rvert,
\end{equation}
following the convention in~\cite{davidLongDistanceSpin2024}.

Ideally, the electron should remain in a single valley eigenstate during shuttling. However, in regions where the valley splitting \(E_v\!\left(x_{qd}\right)\) becomes small, rapid changes in the local valley eigenbasis and enhanced non-adiabaticity can induce unwanted non-adiabatic transitions and generate population in the excited valley state. In the presence of a valley-dependent modulation on the electron $g$-factor, valley dynamics can imprint additional phases and entangle spin and valley degrees of freedom, thereby leaking the spin state during shuttling. The valley maps used in this work are simulated, and the effect of material disorder is included using the alloy-diffusion model following the statistics reported in~\cite{paqueletwuetzAtomicFluctuationsLifting2022}.

\subsection{Time-dependent spin--valley Hamiltonian}
\label{subsec:hamiltonian}

We follow the spin--valley model of~\cite{davidLongDistanceSpin2024}. The qubit is encoded in the spin state of the electron. The spin and the valley states form a four-level system, and the total system Hamiltonian is defined as:
\begin{equation}
H(t) = H_S + H_V\!\left(x_{qd}(t)\right) + H_{VS}\!\left(x_{qd}(t)\right).
\end{equation}

The spin Hamiltonian corresponds to the Zeeman splitting $E_{z}$ in an external magnetic field $B_z$:
\begin{equation}
H_S = \frac{E_Z}{2}\,\sigma_z,
\qquad
E_Z = \bar g\,\mu_B\,B_z.
\end{equation}
where $\bar g$ is the average Landé g-factor of the electron, $\sigma_z$ is the Pauli operator acting on the spin subspace and $\mu_{B}$ is the Bohr magneton. 

The valley contribution to the Hamiltonian is given by the complex intervalley coupling evaluated at the dot position $x_{qd}(t)$,
\begin{equation}
H_V\!\left(x_{qd}(t)\right) =
\Delta_r\!\left(x_{qd}(t)\right)\tau_x +
\Delta_i\!\left(x_{qd}(t)\right)\tau_y,
\end{equation}
where $\tau_x$ and $\tau_y$ are Pauli operators acting in the valley subspace. The position-dependent complex intervalley coupling $\Delta(x_{qd})$ has recently been mapped experimentally with nanometer resolution via conveyor-mode shuttling of entangled spin pairs in a Si/SiGe device \cite{volmerMappingGfactorsComplex2026}, providing direct microscopic access to the valley landscape entering $H_V$. 

The valley--spin coupling arises from valley-dependent $g$-factors and is given by
\begin{equation}
H_{VS}\!\left(x_{qd}(t)\right) =
-\,\kappa_z\,\tilde{\tau}_z\!\left(x_{qd}(t)\right)\otimes\sigma_z,
\qquad
\kappa_z = \frac{1}{4}\,\frac{\delta g}{g}\,E_Z,
\end{equation}
where the rotated valley Pauli operator defines the local valley quantization axis:

\begin{equation}
\tilde{\tau}_z\!\left(x_{qd}(t)\right)
=
\frac{
\Delta_r\!\left(x_{qd}(t)\right)\tau_x +
\Delta_i\!\left(x_{qd}(t)\right)\tau_y
}{
\left|\Delta\!\left(x_{qd}(t)\right)\right|
}.
\end{equation}
\subsection{Open-system dynamics}
\label{subsec:open_system}

We model the open-system dynamics during shuttling using a Markovian Lindblad master equation with a valley relaxation time $T_{1,v}$:
\begin{equation}
\begin{aligned}
\dot{\rho}(t)
&= -\frac{i}{\hbar}\Bigl[H(x_{qd}(t)),\rho(t)\Bigr] \\
&\quad + \frac{1}{T_{1,v}}\,
\mathcal{D}\!\left[\tilde{\tau}_-\!\left(x_{qd}(t)\right)\right]\!\big(\rho(t)\big)
\end{aligned}
\end{equation}

The Lindblad dissipator is defined as
\begin{equation}
\mathcal{D}[L](\rho)=L\rho L^\dagger-\frac{1}{2}\left\{L^\dagger L,\rho\right\}.
\end{equation}
Here $\tilde{\tau}_-\!\left(x_{qd}(t)\right)$ denotes the valley relaxation operator expressed in the local valley eigenbasis at position $x_{qd}(t)$. In the following simulations, we assume that the spin dephasing is not relevant for the timescales of the experiment and assume a long valley lifetime $T_{1,v}$ of 1 ms.

Following Langrock \emph{et al.}~\cite{langrockBlueprintScalableSpin2023}, we assume that the electron remains in the orbital ground state during shuttling. This is supported by the orbital splitting $E_{\mathrm{orb}}(t)=E_1(t)-E_0(t)$, which remains above $1\,\mathrm{meV}$ along the entire trajectory, as shown in Appendix Fig.~\ref{fig:mapping_validation}.

Definitions of the fidelity and spin-purity metrics used to quantify shuttling performance are provided in Appendix~\ref{app:metrics}.

\section{Cryogenic Signal Generator}
\label{sec:signal_generator}

\subsection{Design goals and interface to the qubit}
\label{subsec:design_goals}

\begin{figure}[h!]
  \centering
  \includegraphics[width=0.9\linewidth]{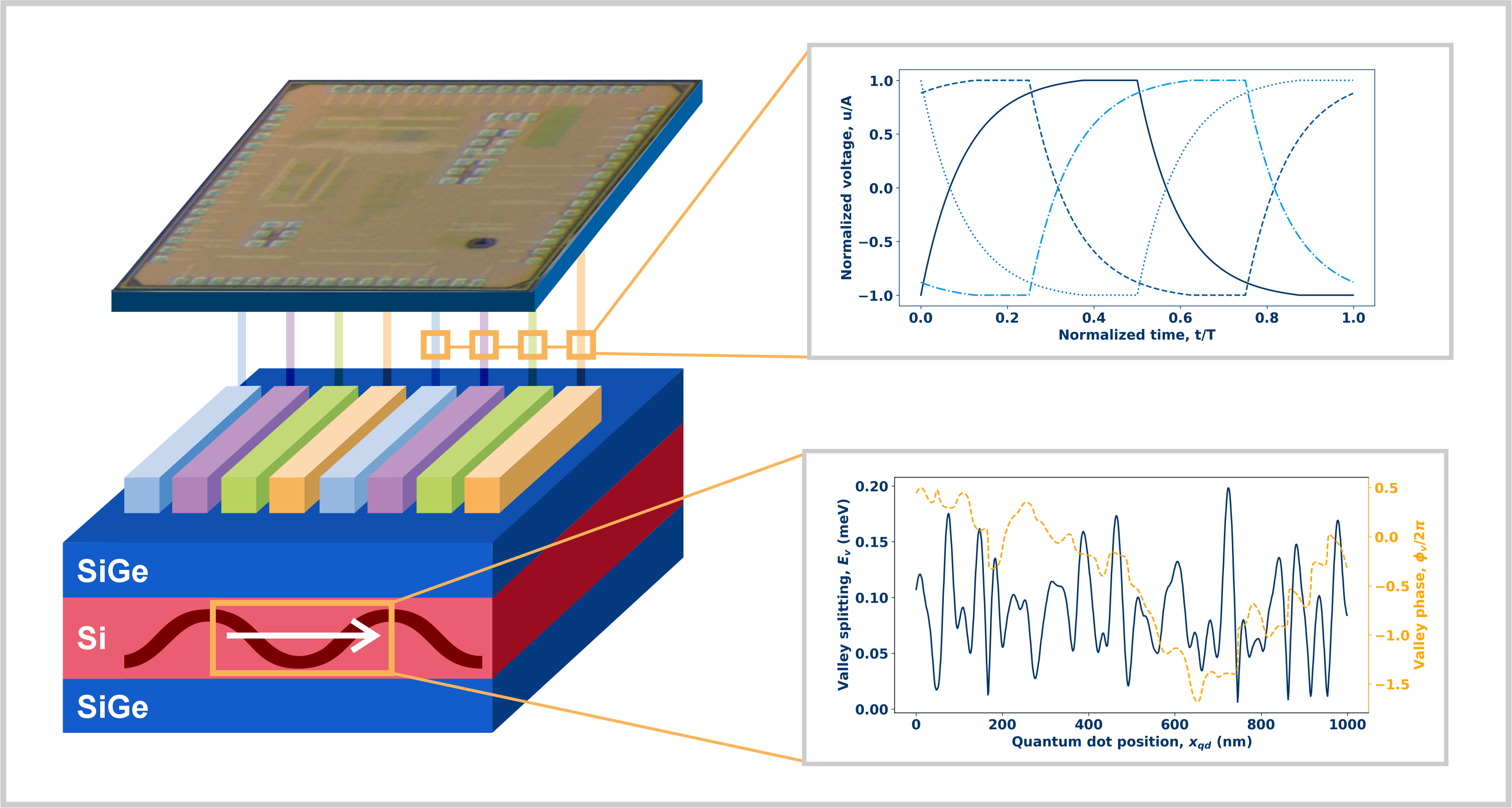}
\caption{Schematic of the integrated cryogenic shuttling signal generator and a Si/SiGe shuttling channel. Four phase-shifted periodic gate voltages (indicated) drive overlapping electrostatic potentials that convey a quantum dot (QD) that confines and transports an electron spin qubit. The plot overlay shows the valley landscape along an illustrative $1~\mu\mathrm{m}$ segment of the shuttling channel: local valley splitting $E_v(x_{\mathrm{qd}})=2|\Delta(x_{\mathrm{qd}})|$ (blue) extracted from a synthetic, disorder-informed valley map, and the corresponding spatial variation of the imaginary part $\Delta_i(x_{\mathrm{qd}})$ (yellow).}

  \label{fig:shuttler}
\end{figure}

The cryogenic signal generator presented in this work is designed for operation at the millikelvin stage, where power dissipation, circuit area, and wiring density are strongly constrained. The circuit generates four phase-shifted periodic gate-voltage waveforms to realize conveyor-mode shuttling. We target wiring schemes as in the SpinBus architecture~\cite{kunneSpinBusArchitectureScaling2024}, where every fourth shuttling gate is electrically connected, reducing the number of independent control signals and enabling a four-phase traveling-wave drive. Fig.~\ref{fig:shuttler} summarizes the resulting interface between the integrated circuit, the shuttling gates, and the valley landscape sampled during transport.

Assuming a gate pitch $l_{\mathrm{pitch}}=100~\mathrm{nm}$ and a four-phase connection scheme, one signal period advances the traveling potential by $4l_{\mathrm{pitch}}$. Therefore, for a targeted average shuttling velocity $v_{\mathrm{avg}}$, the required waveform frequency is
\begin{equation}
f=\frac{v_{\mathrm{avg}}}{4\,l_{\mathrm{pitch}}}.
\label{eq:shuttling_frequency}
\end{equation}

The signal generator enables period-wise programmability of the waveform shape while keeping the period $T=1/f$ fixed, providing a hardware control parameter for shaping the velocity profile within each period without changing the average phase-advance rate set by $f$. Our parametrization requires only one configuration word per signal period and does not require streaming data to the chip during shuttling thanks to an on-chip memory. For example, a shuttling distance of $10~\mu\mathrm{m}$ in a four-phase wiring scheme (translation per period $=4l_{\mathrm{pitch}}$) with $l_{\mathrm{pitch}}=100~\mathrm{nm}$ requires the optimization of $10~\mu\mathrm{m}/(4\cdot100~\mathrm{nm})=25$ configuration words. The fixed clocking and fixed period further simplify timing constraints for the cryogenic electronics.

\subsection{Circuit architecture and control}
\label{subsec:circuit}

\begin{figure}[h!]
  \centering
  \includegraphics[width=0.8\linewidth]{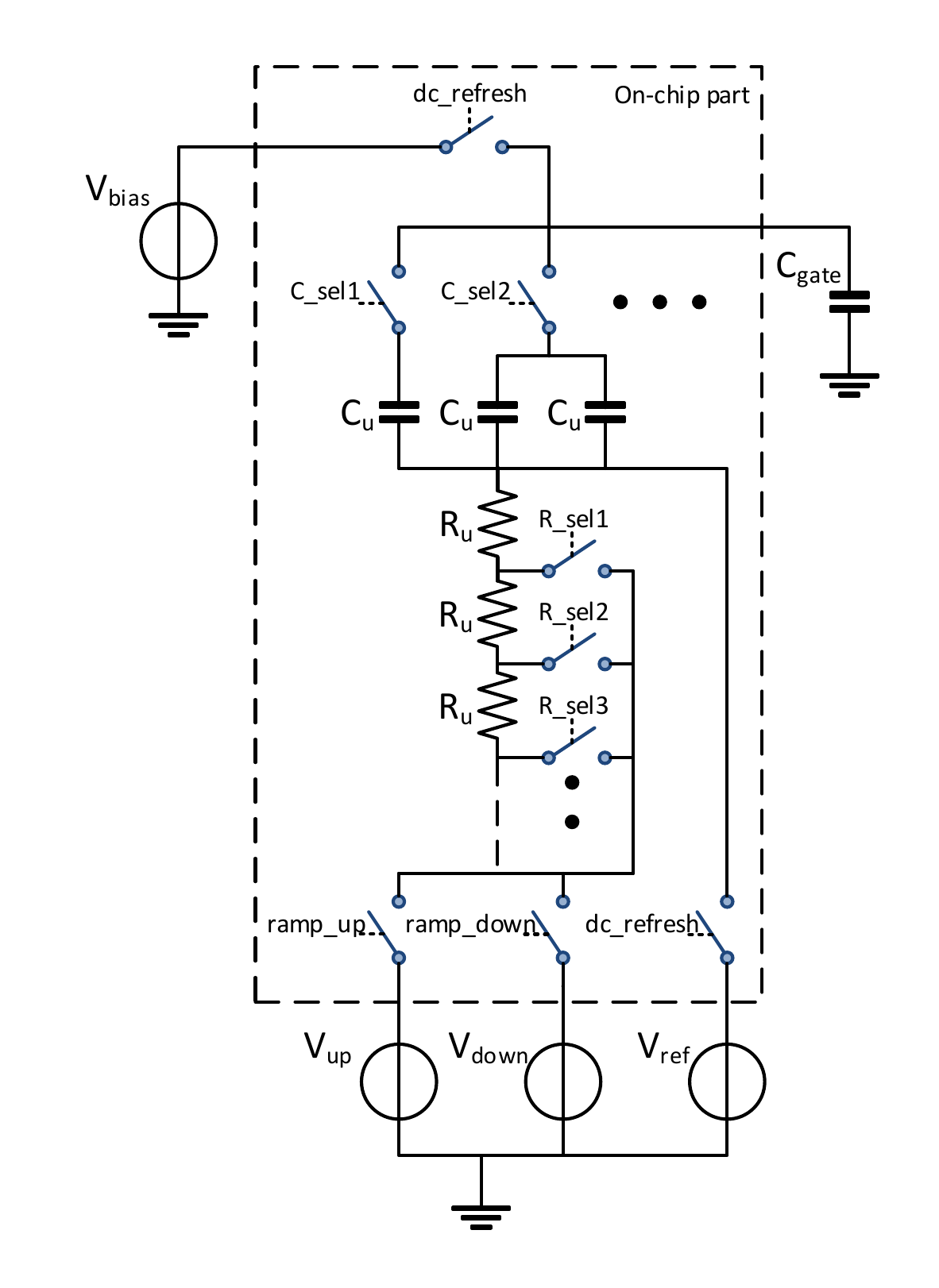}
  \caption{Schematic of the programmable RC network driving the gate capacitance $C_{\mathrm{gate}}$. A selectable capacitor bank ($C_u$) and resistor ladder ($R_u$) set the ramp time constant $\tau$, while control switches enable DC refresh and controlled ramp generation using $V_{\mathrm{bias}}$, $V_{\mathrm{ref}}$, $V_{\mathrm{up}}$, and $V_{\mathrm{down}}$.}
  \label{fig:schematic}
\end{figure}

Fig.~\ref{fig:schematic} shows a single signal-phase generator based on a programmable resistor–capacitor (RC) network. The circuit produces a periodic gate-voltage waveform by alternately charging and discharging the effective capacitive load, generating controlled exponential rise and fall ramps. The analog supply voltages $V_{\mathrm{up}}$ and $V_{\mathrm{down}}$ are supplied from room temperature and define the upper and lower charging rails of the ramp generator. Two additional room-temperature bias inputs, $V_{\mathrm{bias}}$ and $V_{\mathrm{ref}}$, reset the capacitor to a defined voltage during initialization prior to shuttling, while the digital control logic is powered by $V_{dd}$ and driven by an external clock.

Each period consists of a ramp-up phase and a ramp-down phase. Because four phase-shifted shuttling signals are produced, the period is discretized into four clock steps, giving $f_{\mathrm{clk}} = 4f$. For typical operating points with $v_{\mathrm{avg}}=4$--$20~\mathrm{m\,s^{-1}}$ (Eq.~\eqref{eq:shuttling_frequency}), this corresponds to clock frequencies of $40$--$200~\mathrm{MHz}$. By biasing the capacitor to $V_{\mathrm{bias}}$, the waveform is generated around a DC level set by $V_{\mathrm{bias}}$, enabling symmetric voltage swings $(V_{\mathrm{bias}}-A,\,V_{\mathrm{bias}}+A)$ rather than $(0,\,2A)$ for a given amplitude $A$.

The elements $C_u$ form a digitally selectable capacitor bank and $R_u$ a digitally selectable resistor ladder. Together with $C_{\mathrm{gate}}$, they define the RC time constant shaping the rising and falling ramps. The resistor ladder provides the primary control knob for waveform-shape optimization. A configuration word is applied once per period by the state machine, selecting one of four discrete, hardware-realizable resistor-ladder configurations. During shuttling, the DC-refresh switches remain open, while the \texttt{ramp\_up} and \texttt{ramp\_down} switches alternately realize the charge and discharge phases. Further circuit details are given in Appendix~\ref{app:circuit}.

\subsection{Integrated circuit simulation}
\label{subsec:ic_simulation}

The cryogenic shuttling-signal generator was designed in a 22\,nm CMOS fully depleted silicon-on-insulator (FDSOI) technology and simulated in \emph{Cadence Spectre} using an extracted netlist of the integrated circuit design. To approximate cryogenic operation, the standard room-temperature transistor models of the process design kit (PDK) are replaced, where available, by a cryogenic FDSOI transistor model library proposed by Chava \emph{et al.}~\cite{chavaEvaluationCryogenicModel}. 

The electrode waveforms $V_e^{(k)}(t)$ are obtained from noise-enabled transient simulations. Because validated cryogenic noise models are unavailable, the resulting noise levels should be interpreted as an approximate proxy for millikelvin operation.

Further details of the transient Spectre simulations are provided in Appendix~\ref{app:simulation_details}.

\subsection{Mapping gate voltages to the qubit trajectory}
\label{subsec:qubit_position}

The effective RC constant can be updated at the start of each period by selecting one of a finite set of resistor settings (four values
in this work). Consequently, the dimensionless RC time constant $\tau(t)$ is piecewise constant over successive signal periods,
\begin{equation}
\tau(t)=\frac{2\pi R_p C}{T},
\qquad pT \le t < (p+1)T,\quad p=0,1,2,\dots
\label{eq:tau}
\end{equation}
where $R_p$ denotes the resistor value selected for period $p$.

The integrated circuit produces four phase-shifted gate voltages. Each electrode $e\in\{1,\dots,4\}$ is associated with a fixed phase offset $\phi_e$,
\begin{equation}
\phi_e = \bigl[0.25,\,0.75,\,1.25,\,1.75\bigr]\pi,
\end{equation}
and a phase
\begin{equation}
\psi_e(t) = \left(\tfrac{2\pi}{T}t + \phi_e\right) \bmod 2\pi.
\end{equation}

The shuttling signal consists of an exponentially rising and falling waveform within each period, which can be analytically approximated by

{\small
\begin{equation}
\hat{V}_e(t) =
\begin{cases}
-A + 2A\!\left(1 - e^{-\psi_e(t)/\tau(t)}\right), 
& 0 \le \psi_e(t) < \pi \\[6pt]
\;\;\;A - 2A\!\left(1 - e^{-(\psi_e(t)-\pi)/\tau(t)}\right), 
& \pi \le \psi_e(t) < 2\pi.
\end{cases}
\label{eq:ramps}
\end{equation}
}

\begin{figure}[h!]
  \centering
  \includegraphics[width=0.9\linewidth]{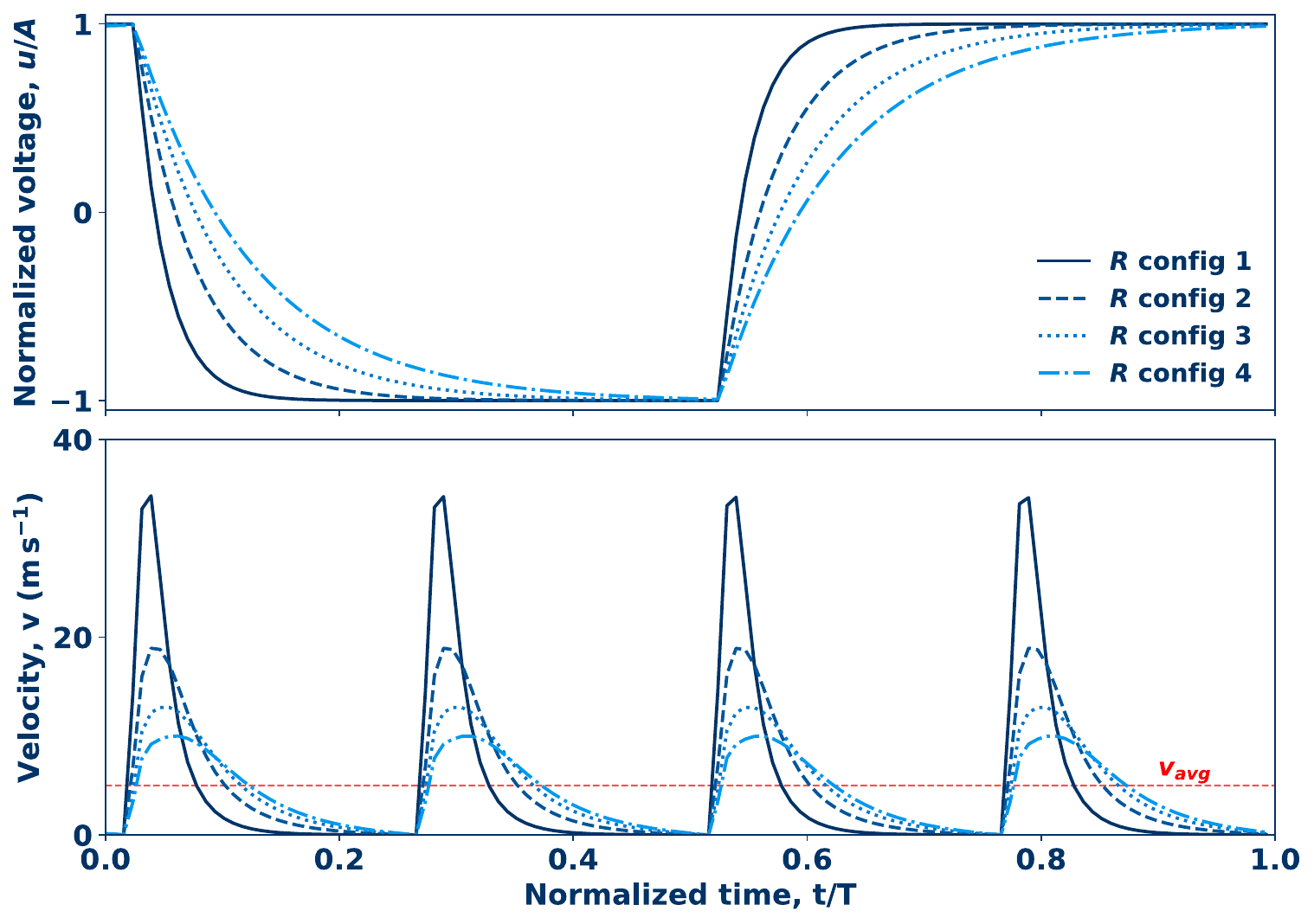}
    \caption{Analytical waveforms illustrating the effect of discrete resistor-ladder settings on ramp shaping and the resulting shuttling-velocity modulation (variations exaggerated for clarity). \textbf{Top:} one gate voltage phase $\hat{V}_e(t)$ normalized to the amplitude $A$, plotted versus normalized time $t/T$ for four discrete RC configurations (different time constants $\tau$; cf. Eqs.~\eqref{eq:tau} and \eqref{eq:ramps}). \textbf{Bottom:} corresponding shuttling velocity $v(t)=\dot{x}_{\mathrm{qd}}(t)$ extracted from the qubit trajectory $x_{\mathrm{qd}}(t)$. Smaller $\tau$ produces steeper ramps and higher velocity peaks, while larger $\tau$ yields smoother ramps and reduced peak velocity. The horizontal line indicates the average velocity $v_{\mathrm{avg}}$.}
  \label{fig:rc_variations}
\end{figure}

This analytical model illustrates how the discrete resistor-ladder settings shape the ramps through the time constant $\tau(t)$ (Eq.~\eqref{eq:tau}). As shown in Fig.~\ref{fig:rc_variations}, varying $\tau$ modifies the exponential rise and fall of the normalized gate voltage $\hat{V}_e(t)/A$ while keeping the period $T$ fixed. 

The trajectory $x_{\mathrm{qd}}(t)$ is obtained by tracking the minimum of the traveling electrostatic potential in \emph{COMSOL}, which captures discretization effects of the gate electrodes but is too costly to repeat for the thousands of waveform realizations required by the optimal-control loop. We therefore use a phasor-projection method as a fast approximation to extract $x_{\mathrm{qd}}(t)$; this approach is defined and validated in Appendix~\ref{app:trajectory}.

The corresponding shuttling velocity is computed by differentiating the trajectory, $v(t)=\dot{x}_{\mathrm{qd}}(t)$. All configurations preserve the same net phase advance per period, and therefore the same average velocity $v_{\mathrm{avg}}$, while redistributing the instantaneous velocity within each period.

\section{Optimization of Control}
\label{sec:optimizer}

We implement controlled velocity modulation during shuttling, closely following the approach previously demonstrated for ideal signal generators~\cite{davidLongDistanceSpin2024}. By reshaping the trajectory $x_{qd}(t)$ and thus the time dependence of $H(t)$, velocity modulation controls both the residence time in regions of small valley splitting $E_v(x_{qd})$, which affects non-adiabatic valley excitation, and the accumulated spin phase induced by the valley-dependent $g$-factor term $H_{VS}(x_{qd})$.

In the optimization loop, the final spin purity $\mathcal{P}_s(t_f)$ serves as the base optimization metric because it directly quantifies residual spin--valley entanglement and is cheaper to evaluate than the full shuttling fidelity. In the noise-free case, the objective is to maximize $\mathcal{P}_s(t_f)$ directly, whereas in the noise-aware case the objective is extended to promote robustness across noise realizations. The shuttling fidelity $\mathcal{F}$ is reported separately as a validation metric to assess phase-stable transport in the rotating frame, consistent with experimental benchmarking. The definitions of the spin-purity and shuttling-fidelity metrics can be found in Appendix~\ref{app:metrics}.

We optimize the discrete RC sequence with a genetic algorithm (GA) based on the Pygad Python package~\cite{gadPyGADIntuitiveGenetic2024}. The shuttling time is divided into periods, and each period is assigned one of four hardware-available resistor-ladder settings. We exploit population-level parallelization to reduce wall-clock time. The GA parameters are listed in Tab.~\ref{tab:params_ga}.

The optimization can be performed either with or without electronic noise. Noise-aware optimization reflects the experimentally relevant operating environment, whereas noise-free optimization is useful for controlled analysis of deterministic effects.

\subsection{Noise-free signal optimization}
With electronic noise disabled, we run the GA to determine a deterministic trajectory. The optimizer evaluates the cost function $J = \mathcal{P}_s(t_f)$. Maximizing $J$ promotes a high final-state purity and therefore minimizes information leakage into the valley states. After each generation, the algorithm updates the sequence of per-period resistance configurations, corresponding to one resistor-ladder setting for each shuttling period.

\subsection{Optimization under cryogenic noise}
In the noise-aware case, we optimize the RC sequence while explicitly incorporating the noise model, using the spin purity defined in Appendix~\ref{app:metrics} as the base optimization metric. For a set of noise realizations, the objective is
\begin{equation}
J_{\mathrm{noise}} = \overline{\mathcal{P}_s}(t_f) - \lambda_{\sigma}\,\sigma_{\mathcal{P}_s}(t_f),
\end{equation}
where $\overline{\mathcal{P}_s}(t_f)$ and $\sigma_{\mathcal{P}_s}(t_f)$ denote the mean and standard deviation of the final spin purity across noise realizations, and $\lambda_{\sigma}$ controls the robustness penalty. The second term suppresses solutions that are overly sensitive to individual noise instances and thereby promotes more uniform performance.

\begin{table}[h!]
\centering
\caption{Parameters for the GA optimization under cryogenic noise}
\label{tab:params_ga}
\begin{tabular}{lc}
\toprule
\textbf{Parameter name} & \textbf{Value}\\
\midrule
Number of generations   & 50     \\
Number of populations   & 12    \\
Number of parents       & 2       \\
Mutation probability    & 0.3     \\
Crossover               & uniform \\
Elitism                 & 1       \\
Parent selection &  tournament \\

\bottomrule
\end{tabular}
\end{table}

\section{Results}
\label{sec:results}

We apply the end-to-end co-simulation workflow of Sec.~\ref{sec:methodology} to hardware-level optimal control of a cryogenic shuttling signal generator. Specifically, we examine how electronic noise limits shuttling fidelity, how this sensitivity depends on the average shuttling velocity, and whether period-wise ramp shaping can improve performance.

\subsection{Noise-induced dispersion of spin purity}
\label{subsec:noise_dispersion_purity}

\begin{figure}[h!]
  \centering
  \includegraphics[width=0.95\linewidth]{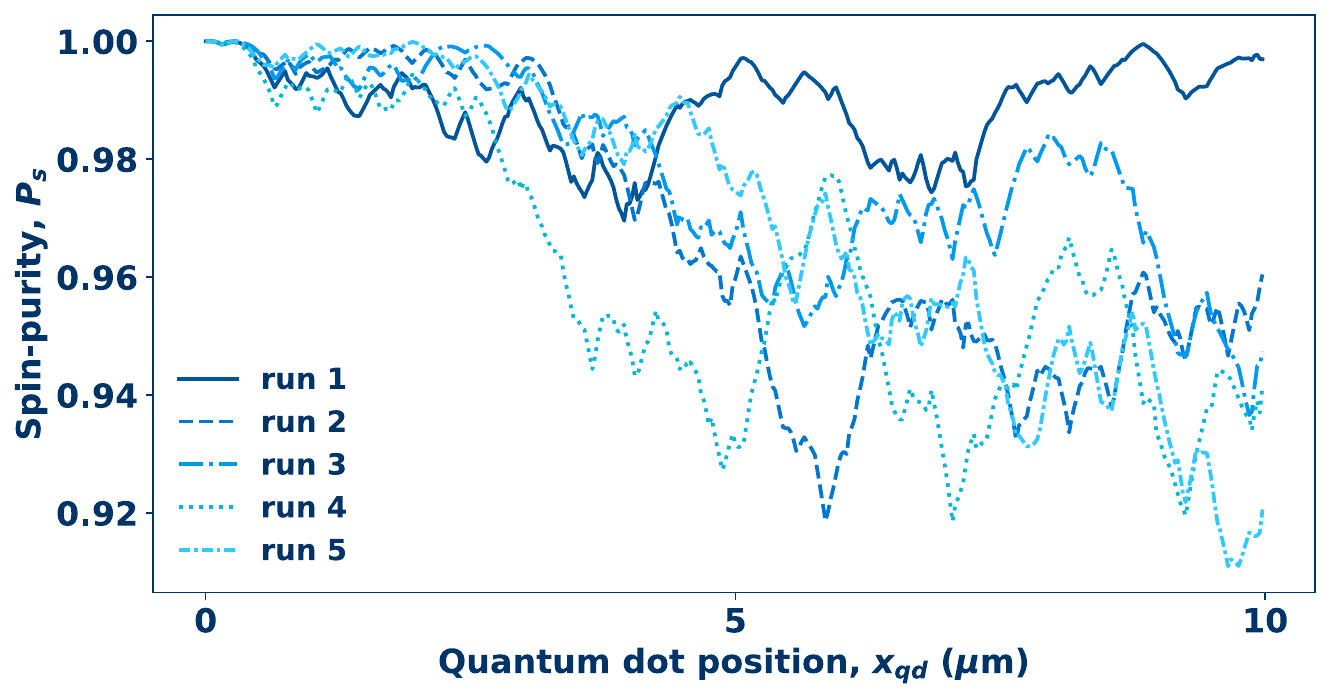}
\caption{Noise-induced purity dispersion at an average shuttling velocity of $v_{\mathrm{avg}}=5\,\mathrm{m\,s^{-1}}$ over a shuttling distance of $10\,\mu\mathrm{m}$ with an ambient temperature of $230~\mathrm{K}$. The dispersion is evaluated over $N_{noise}=50$ independent electronic-noise realizations (shot-to-shot) obtained from transistor-level simulations in \emph{Cadence Spectre} (including noisy circuit components). Five representative purity traces $\mathcal{P}_s^{(k)}(t)$ are shown for clarity. The control sequence is unoptimized (constant resistor configuration).}
  \label{fig:dispersion}
\end{figure}

Fig.~\ref{fig:dispersion} shows representative spin-purity traces $\mathcal{P}_s^{(k)}(t)$ for unoptimized shuttling over $10\,\mu\mathrm{m}$ at an average velocity $v_{\mathrm{avg}}=5\,\mathrm{m\,s^{-1}}$. The dispersion is evaluated over $N_{\mathrm{noise}}=50$ independent shot-to-shot electronic-noise realizations obtained from transistor-level circuit simulations, where each trace corresponds to a distinct noisy waveform. Slow-drift correlations are neglected, so the analysis probes robustness to independent shot-to-shot noise realizations rather than long-timescale drift in the control hardware. The resulting spread in $\mathcal{P}_s^{(k)}(t)$ shows that neglecting electronic noise can lead to overly optimistic performance estimates, highlighting the sensitivity of valley-dependent spin dynamics to noise-induced waveform variations.

\subsection{Influence of shuttling velocity on noise robustness}
\label{subsec:velocity_dispersion_purity}

As described in Subsec.~\ref{subsec:ic_simulation}, the foundry PDK used in this work provides device and noise models only down to approximately $230~\mathrm{K}$. The cryogenic noise behavior of the complete circuit therefore cannot be fully validated in the cryogenic simulation setup, and absolute predictions of circuit-noise-limited shuttling performance remain uncertain. We therefore assess robustness to stochastic shot-to-shot waveform fluctuations using a proxy cryogenic noise model that does not capture slow temporal drift or fully validated millikelvin circuit noise. To gauge the sensitivity of the results to this limitation, electrode waveforms are generated with \emph{Cadence Spectre} under several proxy-noise configurations by varying the PDK temperature parameter and applying a noise-scaling factor in the transient simulations, after which the resulting shot-to-shot dispersion in shuttling fidelity is quantified.

Due to self-heating effects in the electronics \cite{coskunCryogenicSmallsignalNoise2014}, the effective noise temperature is not expected to reach the refrigerator base temperature. For thermal (Johnson--Nyquist) noise, the power spectral density scales linearly with absolute temperature, so reducing the effective temperature from 230~K to 25~K yields a reduction factor of $25/230 \approx 0.1$. To emulate this reduction while keeping the ambient temperature within the PDK-supported range, we scale the noise sources in the noisy transient simulations to $10\%$ of their nominal level. This approximation is plausible because the performance dispersion is predominantly driven by noise components above $\sim 1~\mathrm{MHz}$, as further investigated in Appendix~\ref{app:noise_details}, where thermal noise is expected to dominate and to be strongly reduced at cryogenic temperatures.

To probe sensitivity to uncertain cryogenic noise modeling, we compare three proxy configurations: (i) an ambient-temperature setting at $300~\mathrm{K}$ as a conservative worst case, (ii) the lowest ambient temperature supported by the PDK at $T\approx 230~\mathrm{K}$, and (iii) a lower-noise proxy intended to mimic cryogenic operation with self-heating, implemented by using the same ambient setting $T\approx 230~\mathrm{K}$ together with a noise-scaling factor of $0.1$.

\begin{figure}[h!]
  \centering
  \includegraphics[width=0.95\linewidth]{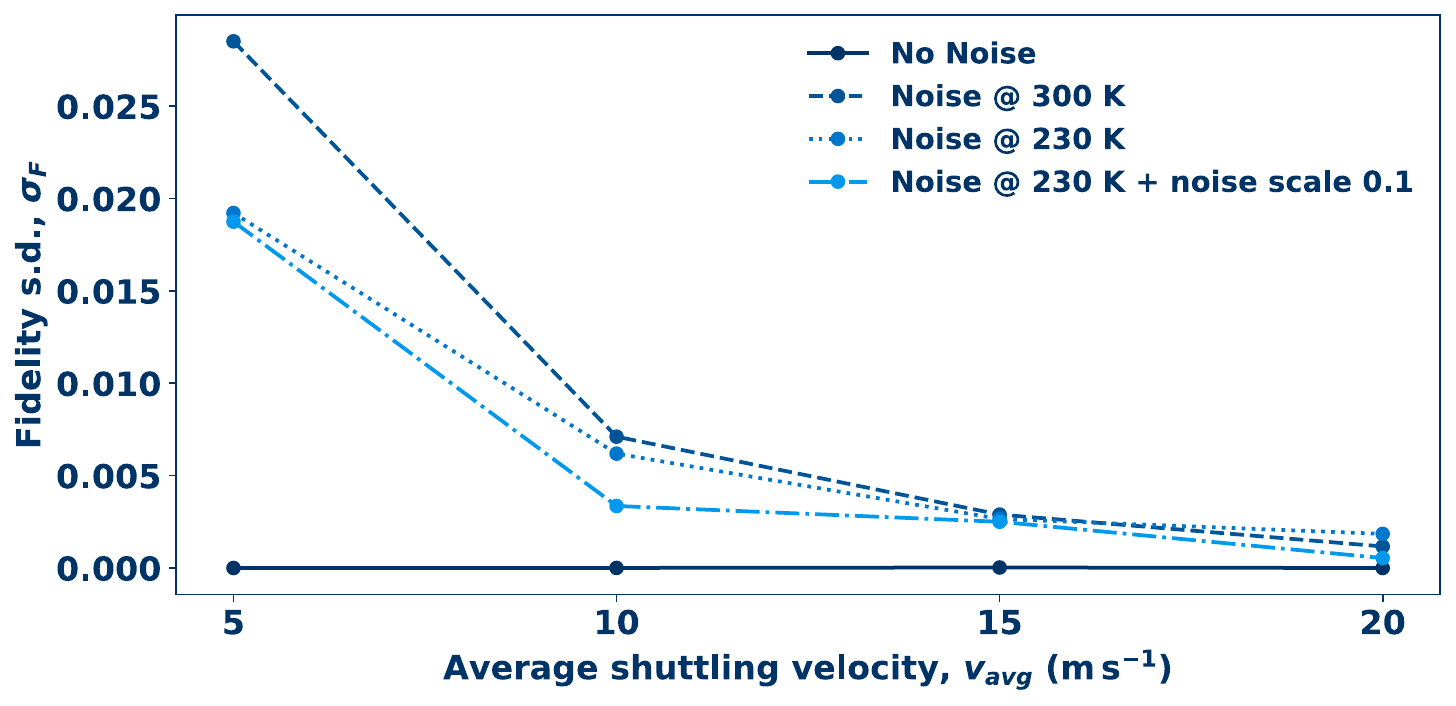}
  \caption{Influence of the average shuttling velocity $v_{\mathrm{avg}}$ on the standard deviation $\sigma_{\mathcal{F}}$ of the shuttling fidelity, evaluated over $N_{\mathrm{noise}}=50$ independent electronic-noise realizations (shot-to-shot) for a shuttling distance of $10~\mu\mathrm{m}$. The shuttling signals are generated with \emph{Cadence Spectre} using a netlist without parasitic extraction. The shuttling sequence is not optimized and is controlled by a fixed resistor sequence. Several noise proxy configurations are compared to explore the sensitivity of $\sigma_{\mathcal{F}}$ to uncertain cryogenic noise modeling, alongside a noise-free reference.}
  \label{fig:speed_noise_relation}
\end{figure}

Fig.~\ref{fig:speed_noise_relation} summarizes the shot-to-shot dispersion of the shuttling fidelity under injected electronic noise. Across all proxy configurations, the dependence on the average shuttling velocity $v_{\mathrm{avg}}$ is the dominant trend: the standard deviation of the fidelity $\sigma_{\mathcal{F}}$ decreases rapidly with increasing $v_{\mathrm{avg}}$ and approaches a saturation regime near $20~\mathrm{m\,s^{-1}}$.

The strong suppression of dispersion at higher shuttling velocities is consistent with a reduced time window for noise-induced phase accumulation and spin--valley entanglement during transport. This picture suggests a motional-narrowing-like mechanism, where faster motion effectively increases the valley population fluctuations, which then approach a steady state with an average occupation. Importantly, the observed saturation above $\sim 20~\mathrm{m\,s^{-1}}$ is in qualitative agreement with recent experimental results by De~Smet \emph{et al.}, who report a pronounced reduction of the phase-flip probability that saturates for shuttling velocities above $\sim 20~\mathrm{m\,s^{-1}}$ (see Fig.~3(d)~\cite{desmetHighfidelitySinglespinShuttling2025}). Although our simulations rely on a proxy-noise model, this qualitative correspondence suggests that increasing $v_{\mathrm{avg}}$ improves noise robustness in long-distance spin transport.

Increasing $v_{\mathrm{avg}}$, however, requires higher shuttling-signal frequencies, which tend to increase instantaneous power dissipation and tighten power constraints on the cryogenic electronics. We therefore select $v_{\mathrm{avg}}=20~\mathrm{m\,s^{-1}}$ as the operating point for subsequent noise-aware optimization, since it lies near the dispersion saturation.

\subsection{Shuttling optimal control}
\label{subsec:optimization_results}

This work next investigates how hardware-embedded shuttling control can (i) compensate deterministic errors arising from valley disorder and (ii) maintain robust performance in the presence of stochastic electronic noise.

\begin{figure}[h!]
  \centering
  \includegraphics[width=0.95\linewidth]{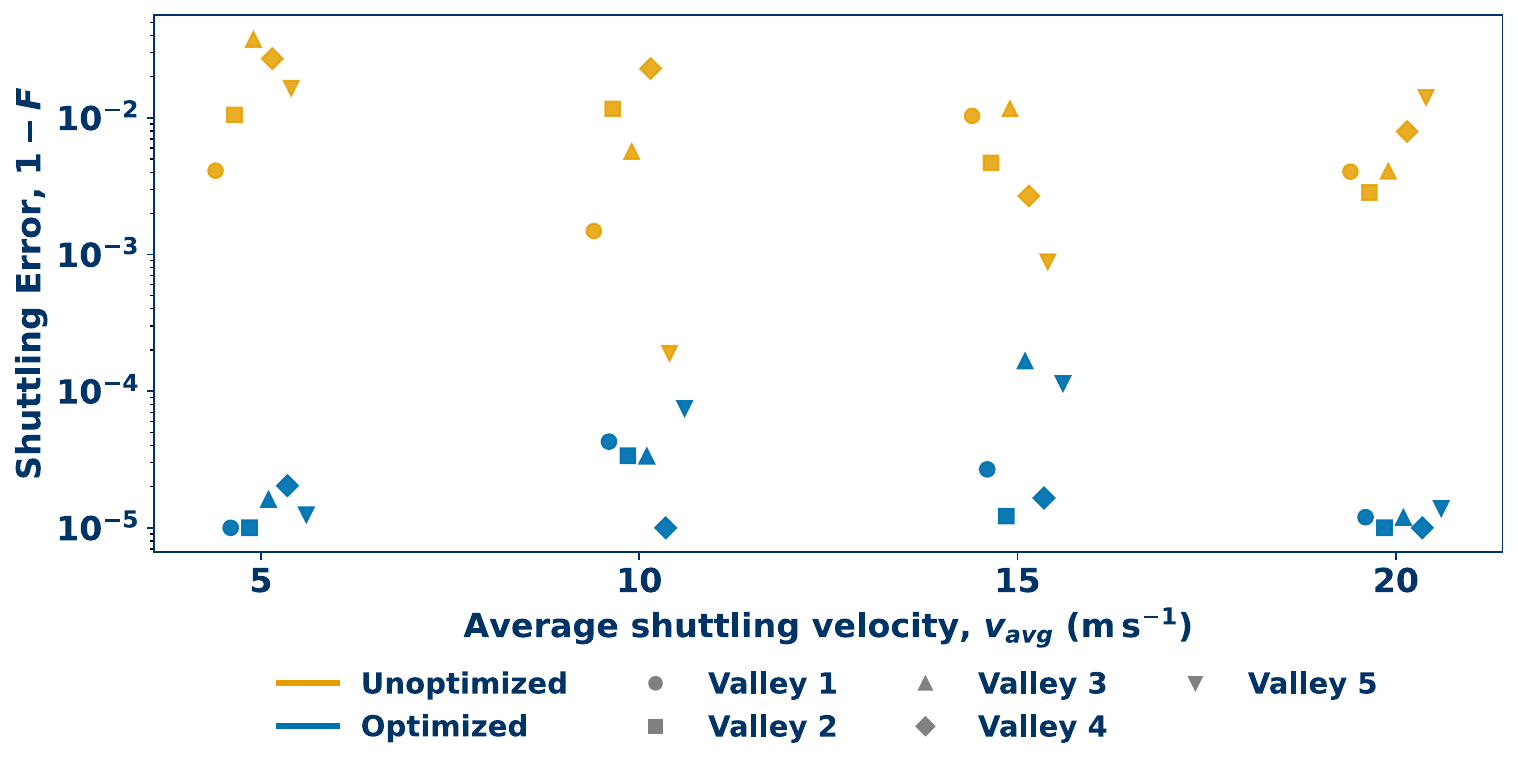}
  \caption{
    Benchmark of hardware-embedded shuttling control against a constant-resistor baseline with electrical noise disabled.
    Markers report the ensemble mean over 50 repeated co-simulation runs for each of five valley maps, for a $10~\mu\mathrm{m}$ shuttle at $v_{\mathrm{avg}}=5$--$20~\mathrm{m\,s^{-1}}$.
  }
  \label{fig:purity_benchmark}
\end{figure}

Fig.~\ref{fig:purity_benchmark} summarizes the shuttling error $1-F$ for four representative average shuttling velocities ($5$--$20~\mathrm{m\,s^{-1}}$). During this initial optimization stage, electrical noise is disabled such that run-to-run variations reflect numerical variability only (typical standard deviations $<10^{-6}$). The genetic algorithm (GA) is terminated once the target threshold $\bar{P}_s(t_f) > 0.9999$ is reached.

Across five valley-disorder realizations, the optimized velocity modulation consistently suppresses the shuttling error relative to the constant-resistor baseline, demonstrating that valley-induced errors can be compensated within the constraints of the control parameterization. In the absence of additional stochastic noise sources, the resulting spin-state fidelities approach levels commonly assumed for fault-tolerant operation.

Higher shuttling velocities substantially reduce the optimization effort. As summarized in Tab.~\ref{tab:complexity} in the Appendix, the GA requires many more generations to reach the target threshold at $v_{\mathrm{avg}} = 10~\mathrm{m\,s^{-1}}$, whereas at $v_{\mathrm{avg}} = 20~\mathrm{m\,s^{-1}}$ convergence is reached within $1$--$4$ generations. Together with the hardware bandwidth constraints and the observed saturation of dispersion suppression at high velocity (Sec.~\ref{subsec:velocity_dispersion_purity}), this motivates selecting $v_{\mathrm{avg}}=20~\mathrm{m\,s^{-1}}$ as the operating point for the subsequent shuttling control.

\begin{figure}[h!]
  \centering
  \includegraphics[width=0.95\linewidth]{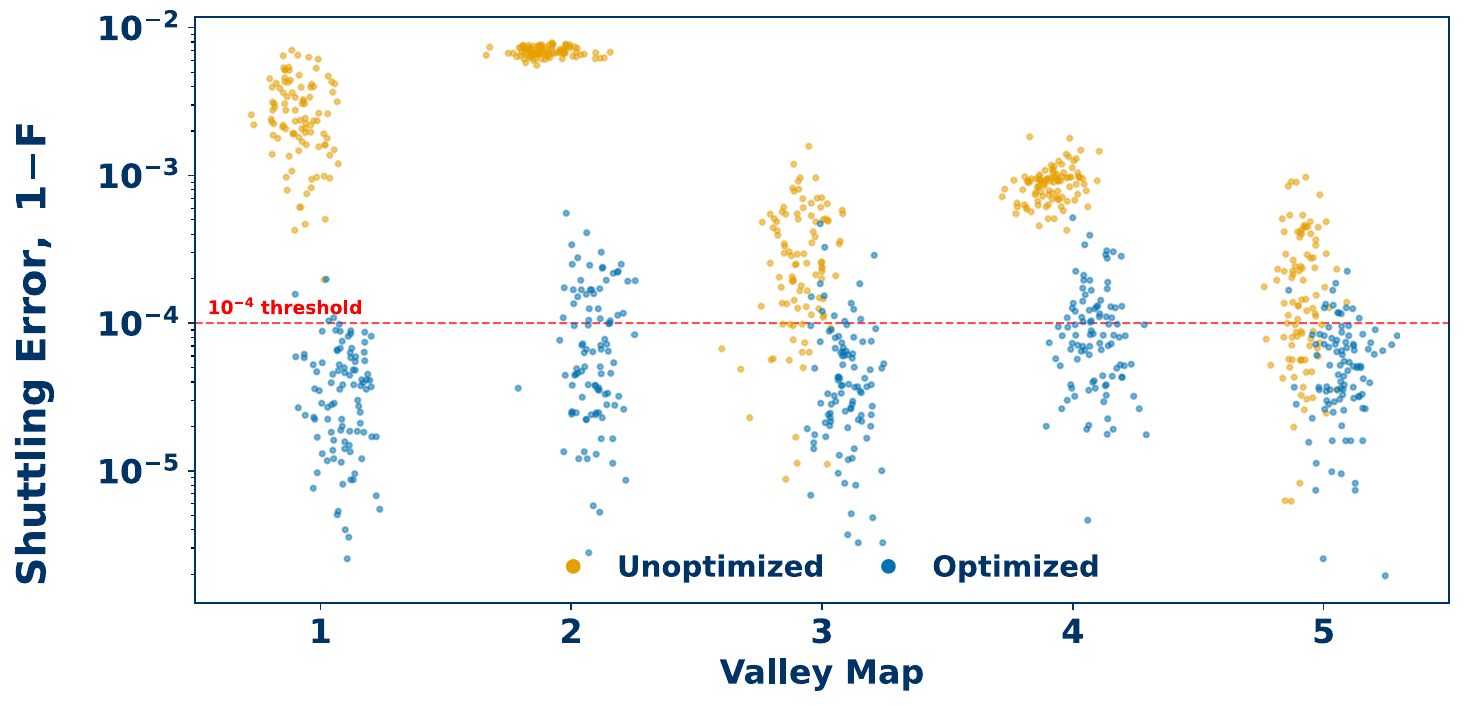}
    \caption{
Shuttling error $1-F$ for five valley-map realizations comparing the unoptimized constant-resistor baseline (orange) with the optimized control (blue) under a cryogenic-noise proxy. All electronic simulations were run at an ambient temperature of $230\,\mathrm{K}$ (within the characterized range of the PDK), and the noise amplitude was reduced by a factor of $10$ using the noise-scale parameter. Each point corresponds to one of $N_{noise} = 100$ independent noise realizations obtained from transient analyses in \emph{Cadence Spectre}. The dashed red line indicates the target threshold $1-F = 10^{-4}$.
}
  \label{fig:fidelity_benchmark_noise}
\end{figure}

The shuttling control is then re-optimized using the GA at $v_{\mathrm{avg}}=20~\mathrm{m\,s^{-1}}$ with electrical transient noise enabled. Fig.~\ref{fig:fidelity_benchmark_noise} compares the optimized resistor sequence with the unoptimized constant-resistor baseline under a cryogenic-noise proxy implemented in \emph{Cadence Spectre}. For a fair paired comparison, the optimized and unoptimized protocols were evaluated for each valley-map realization using the same set of $N_{noise}=100$ independent electronic-noise realizations. The optimized protocol consistently reduces the shuttling error and produces a tighter distribution of fidelities across all valley maps.

At a noise scale of $0.1$, the optimized protocol achieves an average valley-wise fidelity of $99.99 \pm 0.007\%$, compared to $99.78 \pm 0.06\%$ for the unoptimized baseline. Furthermore, $79\%$ of optimized shuttling operations reach the target threshold $1-F = 10^{-4}$ across the five valley maps, compared to $12\%$ for the unoptimized protocol, indicating that under the assumed cryogenic noise conditions the control strategy retains substantial benefit even in the presence of stochastic electronic noise. Notably, these improvements are achieved despite the strongly quantized control space, in which each signal period can only take one of four resistor configurations.

While noise-aware optimal control confirms robust improvements in the fidelity, the dispersion of the excited-valley population remains comparatively large. As quantified by the standard deviation $\sigma_{p_v(t_f)}$ (see Appendix~\ref{app:metrics}), the final excited-valley population $p_v(t_f)$ exhibits only a weak dependence on $v_{\mathrm{avg}}$ within the present model (Tab.~\ref{tab:valley_state}). This suggests that electronic noise induces largely uncontrolled redistribution in the valley state during long-distance transport, such that the present control strategy primarily enhances spin coherence rather than stabilizing a unique valley state.

\begin{table}[h!]
\centering

\caption{Final excited-valley population $p_v(t_f)$ as a function of average shuttling velocity $v_{\mathrm{avg}}$. Values are reported as mean $\pm$ standard deviation over $N_{\mathrm{noise}}=50$ independent electronic-noise realizations for an optimized shuttling sequence, validated using the co-simulation framework.}

 \label{tab:valley_state}
\begin{tabular}{lc}
\toprule
$v_{avg}$ (m\,s$^{-1}$) & $p_v(t_f)$ \\
\midrule
    5  & $0.4405 \pm 0.2925$ \\
    10 & $0.4219 \pm 0.2961$ \\
    15 & $0.5688 \pm 0.2811$ \\
    20 & $0.5775 \pm 0.2561$ \\
\bottomrule
\end{tabular}
\end{table}

\subsection{Power dissipation of the cryogenic shuttling generator}

The analog power consumption is extracted from transient \emph{Cadence Spectre} simulations including device and interconnect parasitics. Tab.~\ref{tab:power_benchmark} summarizes the active analog power dissipation \(P_a\) drawn from the supply rails \(V_{\mathrm{up}}\) and \(V_{\mathrm{down}}\) in Fig.~\ref{fig:schematic} as a function of the average shuttling velocity \(v_{\mathrm{avg}}\).

The observed \(\mu\)W-level dissipation is consistent with predominantly capacitive switching. The effective switched capacitance is set by the series combination of the capacitor bank, parasitic capacitances, and the gate/electrode capacitance, and is therefore expected to be much closer to \(C_{\mathrm{gate}} \approx 1~\mathrm{pF}\) than to the capacitor-bank value of approximately \(10~\mathrm{pF}\). Using the repetition frequency \(f\) based on Eq.~(\ref{eq:shuttling_frequency}), and a peak-to-peak voltage \(V_{\mathrm{pp}} = 200~\mathrm{mV}\), the expected scaling \(P \propto C_{\mathrm{eq}} V_{\mathrm{pp}}^2 f\) is consistent with the trend in Tab.~\ref{tab:power_benchmark}.

The digital switching of the state machine is not included in the analog signal-generation power and is expected to add an overhead that does not scale linearly with the analog power.

\begin{table}[h!]
\centering
\caption{Active analog power dissipation \(P_a\) of the four-phase shuttling signal generator versus average shuttling velocity \(v_{\mathrm{avg}}\). The shuttling duration \(t_{\mathrm{shuttle}}\) corresponds to a qubit cell-to-cell distance of \(10~\mu\mathrm{m}\), assuming \(t_{\mathrm{shuttle}} = 10~\mu\mathrm{m}/v_{\mathrm{avg}}\).}
\label{tab:power_benchmark}
\begin{tabular}{lcc}
\toprule
$v_{\mathrm{avg}}$ (m\,s$^{-1}$) & $P_a$ ($\mu$W) & $t_{\mathrm{shuttle}}$ ($\mu$s)\\
\midrule
5  & 2.6 & 2.00\\
10 & 5.0 & 1.00\\
15 & 7.8 & 0.67\\
20 & 9.8 & 0.50\\
\bottomrule
\end{tabular}
\end{table}

\section{Discussion}
\label{sec:discussion}

Non-deterministic electronic noise is a central limitation for high-fidelity spin shuttling because it introduces shot-to-shot variability in the spin--valley evolution and therefore reduces the average shuttling fidelity. Our results show that hardware-embedded optimization can recover high performance in the presence of circuit noise and spatially varying valley landscapes, underscoring the need to treat control electronics and qubit dynamics as a coupled system.

\subsection{Comparison with prior theoretical and experimental work}
Nagai \emph{et al.}~\cite{nagaiDigitallyControlledConveyorbelt2025} report shuttling fidelities approaching $99.9\%$ at an average velocity of $10~\mathrm{m\,s^{-1}}$ in a theoretical study focused on cryogenic waveform generation. Their analysis assumes an idealized valley environment with a constant valley splitting of $200~\mu\mathrm{eV}$ and does not explicitly include electronic noise from the control circuitry. 
Our work establishes a framework incorporating disorder-informed valley landscapes and transistor-level circuit non-idealities, enabling a hardware-embedded, noise-aware assessment of shuttling performance under cryogenic electronics constraints.

Recent experimental progress by De~Smet \emph{et al.}~\cite{desmetHighfidelitySinglespinShuttling2025} demonstrates high-fidelity single-spin shuttling in silicon using a two-tone conveyor-mode approach. By employing eight sinusoidal control signals, they achieve a shuttling fidelity of $99.54 \pm 0.03\%$ over a distance of $10~\mu\mathrm{m}$, corresponding to a shuttling velocity of about $54~\mathrm{m\,s^{-1}}$. The observed saturation of phase-error probabilities above velocities of roughly $20~\mathrm{m\,s^{-1}}$ is in qualitative agreement with our simulation results. This agreement is consistent with the interpretation that increasing shuttling velocity reduces the effective interaction time with low-frequency noise sources and thereby improves robustness against electronic noise.

Losert \emph{et al.}~\cite{losertEffectsAlloyDisorder2025} provide a useful outlook for the scenario in which alloy disorder effectively randomizes the valley degree of freedom: they show that high-fidelity single-qubit operations can still be achieved through pulse-level control. This suggests a potential route to mitigating the valley-population randomization observed in our shuttling simulations.

\subsection{System-level scalability}
Cryogenic waveform generation reduces the number of high-bandwidth analog lines routed from room temperature, alleviating cryostat cabling constraints and reducing analog fan-out \cite{geckControlElectronicsSemiconductor2019}. Sharing clock distribution, digital interfaces, and supply rails across many signal generators further reduces wiring overhead and the associated thermal links to warmer cryostat stages.

The circuit in this work also enables a calibration-and-replay workflow: for each shuttling lane, the period-wise RC configuration sequence can be calibrated in a closed loop, stored in on-chip memory, and replayed by an internal state machine. This avoids continuous streaming of high-rate waveform samples into the cryostat during shuttling. By comparison, many DAC-based shuttling demonstrations require streaming of sample data, even when the DACs are operated cryogenically \cite{meyerCryogenicEvaluationDigitaltoAnalog2025,siebererCryogenicDigitaltoAnalogConverter2024,stuartChipIntegratedVoltageSources2019}.

The active analog power dissipation per qubit during shuttling is non-negligible at the millikelvin stage. However, the duty cycle estimated in Appendix~\ref{app:duty_cycle} reduces the average dissipation and sets an upper bound on the achievable time-multiplexing factor. If the mK power budget remains limiting, mitigation options include operating qubits at warmer cryostat stages \cite{vandersypenInterfacingSpinQubits2017} or partitioning the shuttling signal chain \cite{nagaiDigitallyControlledConveyorbelt2025} to move the most power-intensive waveform synthesis away from the mixing chamber.

\section{Conclusion}
\label{sec:conclusion}

We have demonstrated valley-aware high-fidelity electron shuttling with integrated cryogenic control electronics using a co-simulation framework that combines valley disorder, circuit noise, and hardware-level optimal control. By coupling transistor-level circuit simulations to spin--valley dynamics, this work provides an experimentally relevant assessment of shuttling fidelity. We find that electronic noise reduces the average shuttling fidelity and introduces substantial statistical dispersion, underscoring the importance of noise-aware hardware optimization. Because the present study relies on proxy cryogenic noise modeling, the reported noise robustness should be understood as evidence of feasibility rather than a fully predictive millikelvin hardware assessment.

Our results identify ultra-low-power cryogenic CMOS electronics as a promising platform for shuttling control, as the circuitry can directly implement velocity-modulation-based optimal control with active analog power consumption during shuttling at levels compatible with millikelvin operation. Overall, these findings support integrated cryogenic electronics as a promising route toward scalable shuttling-based spin-qubit architectures with embedded valley-aware control over micrometer-scale transport distances.

\section*{Acknowledgments}
\label{sec:acknowledgements}

The authors thank Prof.~Hendrik Bluhm (PGI-11), Akshay Menon Pazhedath (PGI-8) and Lea Schreckenberg (PGI-4) for valuable discussions that contributed to the interpretation of the results. They also thank Ivonne Pajunk (PGI-4) for assistance with the graphics and Mario Schl\"osser (PGI-4) for helpful feedback on the manuscript.

The authors further acknowledge funding from Germany's Excellence Strategy through the Cluster of Excellence Matter and Light for Quantum Computing (ML4Q2), EXC 2004/2 – 390534769; from the Federal Ministry of Education and Research (BMBF) within the framework programme ``Quantum technologies -- from basic research to market'' (project QSolid, Grant No.~13N16149); and from the European flagship project OpenSuperQPlus100 (Grant Agreement No.~101113946, HORIZON-CL4-2022-QUANTUM-01-SGA).

\section*{Competing interests}
P.D.R., L.D., and L.G. are named as inventors on a patent application related to the integrated cryogenic shuttling signal generator presented in this work. P.D.R., L.G., and S.v.W. are founders and shareholders of IceCirc GmbH. The remaining authors declare no competing interests.

\section*{Data availability}
The data that support the findings of this study are available from the corresponding authors upon request.

\section*{Author contributions}
P.D.R. developed the co-simulation framework and the concept of controlled velocity modulation using discrete hardware settings. A.D. developed the qubit model and the Python library for the qubit dynamics. L.D. designed the integrated circuit. N.C. developed the optimizer. N.C. and P.D.R. performed the simulations and optimization studies. L.G., S.v.W., and F.M. provided scientific guidance and supervision.

\appendix

\section{Electronic circuit details}\label{app:circuit}
\begin{figure}
  \centering
  \includegraphics[width=0.95\linewidth]{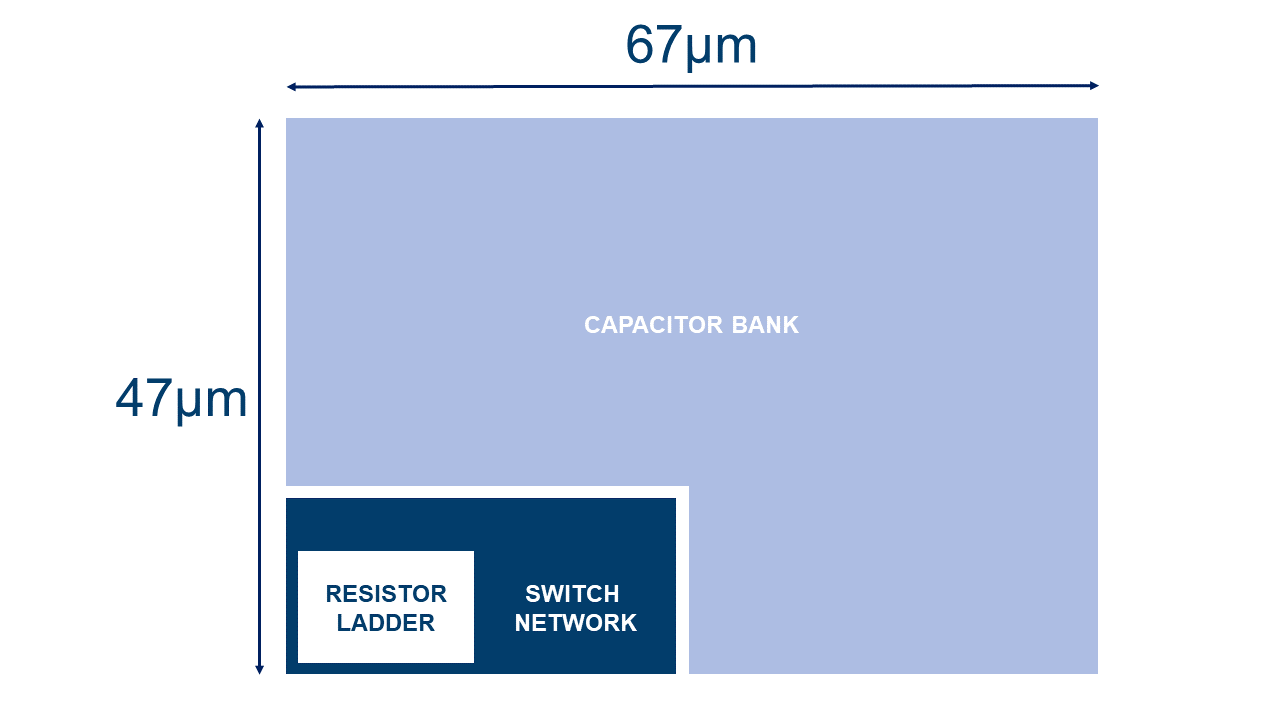}
  \caption{Simplified layout of a single shuttling-signal phase generator in a $22\,\mathrm{nm}$ FDSOI CMOS technology. The four-phase shuttling generator consists of four identical phase cells. The configurable capacitor bank ($C_u$ in Fig.~\ref{fig:schematic}) dominates the area at $\approx 2600~\mu\mathrm{m}^2$ per phase generation, followed by the switch network ($\approx 340~\mu\mathrm{m}^2$) and resistor ladder ($R_u$, $\approx 140~\mu\mathrm{m}^2$). The total area is $\approx 3080~\mu\mathrm{m}^2$ per phase cell. The shared digital logic is excluded because its area does not scale with the number of output phases.}

  \label{fig:layout}
\end{figure}

The layout shown in Fig.~\ref{fig:layout} is replicated four times to generate four \(90^\circ\) phase-shifted shuttling signals. The capacitor bank dominates the footprint: an effective capacitance of order \(10~\mathrm{pF}\) is used to reduce sensitivity to process variations and to suppress \(kT/C\)-type reset noise during initialization (\(\sigma_V \propto \sqrt{kT/C}\)). The capacitor bank was over-dimensioned, providing a large safety margin and reducing the impact of noise. We assume that, in future, the capacitance could be reduced by one order of magnitude. The actual tolerable noise limit is unknown and must be determined experimentally.

The four-phase signal generator occupies \(\approx 12{,}300~\mu\mathrm{m}^2\), far exceeding the \(\approx 100~\mu\mathrm{m}^2\) SpinBus qubit-cell target \cite{kunneSpinBusArchitectureScaling2024}; per-cell integration would therefore require \(\gtrsim 100\times\) area reduction. While switch/transistor area can be reduced by scaling to smaller technology nodes, the capacitor bank dominates the footprint, so higher capacitance density is the primary lever. Deep-trench and related pore/trench decoupling capacitors can provide substantially higher density than planar MOS options, with reported footprint reductions of order \(\sim 8\times\) for comparable capacitance \cite{peiNovelLowcostDeep2008,roozeboomALDOptionsSiintegrated2007}.

At higher shuttling velocity, the duty cycle can decrease (e.g., to \(\sim 1/5\)), implying an idealized time-multiplexing upper bound of \(\sim 5\) shuttling lanes per signal generator (cf.\ Appendix~\ref{app:duty_cycle}).

\section{Trajectory extraction + COMSOL validation}\label{app:trajectory}

To extract the traveling-wave phase from the four gate voltages, we construct the complex phasor
\begin{equation}
z(t) = \sum_{e=1}^{4} V_e(t)\, e^{-j\phi_e},
\label{eq:z}
\end{equation}
and compute the continuous phase evolution as the unwrapped argument
\begin{equation}
\theta(t) = \operatorname{unwrap}\!\big(\arg z(t)\big).
\label{eq:theta}
\end{equation}

The quantum-dot position is obtained by mapping the accumulated phase $\theta(t)$ to the spatial period $L$ of the traveling potential:
\begin{equation}
x_{qd}(t) = \frac{L}{2\pi}\,\theta(t),
\qquad
x_{qd}(0)=0.
\label{eq:phasor}
\end{equation}
Here $L$ is set by the electrode pitch and the connection periodicity (e.g., $L=4\,l_{\mathrm{pitch}}$ for a four-phase connection scheme).

\begin{figure}
  \centering
  \includegraphics[width=0.9\linewidth]{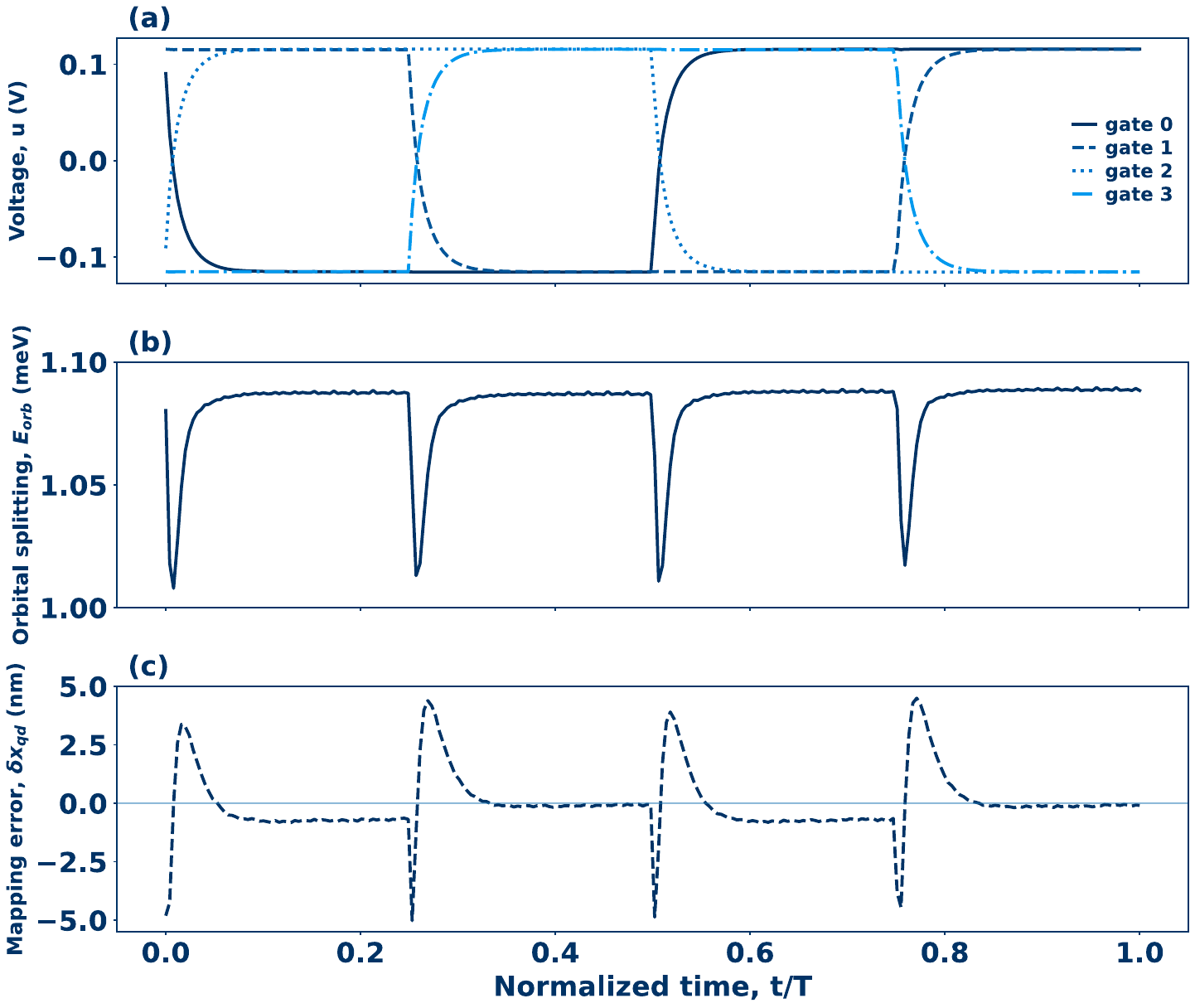}

\caption{Validation of orbital confinement and trajectory extraction used in the co-simulation workflow. \textbf{(a)} Representative four-phase gate-voltage waveforms $V_e(t)$ that generate the traveling electrostatic potential during one normalized period $t/T$. \textbf{(b)} Corresponding orbital energy splitting along the shuttling trajectory, confirming that the dot remains strongly confined. \textbf{(c)} Dot-position mapping error $\delta x(t)$ between the fast phasor-based extraction of $x_{\mathrm{qd}}(t)$ (Eqs.~\ref{eq:phasor}) and a reference mapping obtained by tracking the minimum of the electrostatic potential in a \emph{COMSOL} simulation, showing only few-nanometer deviations with peaks during rapid waveform transitions.}

\label{fig:mapping_validation}
\end{figure}

The phasor-based extraction agrees with the \emph{COMSOL} reference to within a few nanometers, as shown in Fig.~\ref{fig:mapping_validation}, while reducing the computation time by approximately three orders of magnitude.

\section{Metrics}\label{app:metrics}
Unless stated otherwise, all reported means and standard deviations are computed over an
ensemble of $N_{\mathrm{noise}}$ electronic-noise realizations. Each realization
$k=1,\ldots,N_{\mathrm{noise}}$ corresponds to an independent noise-enabled transient circuit
simulation, which produces a unique noisy electrode waveform
$V_e^{(k)}(t)$. This waveform is propagated through the electrode-to-trajectory mapping to
obtain $x_{\mathrm{qd}}^{(k)}(t)$ and is subsequently used in the spin--valley dynamics,
yielding a final joint state $\rho^{(k)}(t_f)$. Throughout the results presented here, the
valley-disorder realization (valley map) is held fixed unless explicitly noted.

\subsection{Spin purity}
Let $t_f$ be the final time of the shuttling sequence. We compute the spin purity over a set of noise seeds where each trajectory returns a noise-influenced final state $\rho^{(k)}(t_{f})$. The average density matrix over the set of trajectories is then computed as:
\begin{equation}
    \bar{\rho}(t_f) = \frac{1}{N_{\text{noise}}} \sum_{k=1}^{N_{\text{noise}}} \rho^{(k)}(t_f)
\end{equation}
To obtain the spin state, we compute the partial trace over the valley of the average density matrix $\bar{\rho}(t_f)$: 
\begin{equation}
    \bar{\rho}_s(t_f) = \mathrm{Tr}_V\!\left[\bar{\rho}(t_f)\right]
\end{equation}
The final spin purity is computed from the noise-averaged density matrix as:
\begin{equation}
    \bar{\mathcal{P}}_s(t_f) = \mathrm{Tr}\!\left[\bar{\rho}_s(t_f)^2\right]
\end{equation}
A reduction of $\bar{\mathcal{P}}_s(t)$ indicates residual spin--valley entanglement and incoherent mixing after tracing out the valley subsystem.

\subsection{Excited valley state population}
The quantity $p_v(t_f)$ measures the residual occupation of the excited valley state at the final quantum dot position. Let
\begin{align}
\rho_v(t_f)
&= \mathrm{Tr}_s\!\left[\rho(t_f)\right]
\end{align}
be the reduced valley density matrix, and $\ket{e(x_{\mathrm{qd}}(t_f))}$ the local excited-valley eigenstate. The excited-valley population is defined as
\begin{align}
p_v(t_f)
&= \bra{e(x_{\mathrm{qd}}(t_f))}
\,\rho_v(t_f)\,
\ket{e(x_{\mathrm{qd}}(t_f))}.
\end{align}

A nonzero population in the excited valley state indicates a non-adiabatic evolution in the valley subspace leading to leakage and dephasing of the spin state through spin--valley coupling.

\subsection{Shuttling fidelity}

We define the fidelity \(\mathcal{F}\) as the mean state fidelity with respect to a reference (“center”) state on the Bloch sphere. The center state is obtained from a test ensemble of 50 final states generated using previously unseen noise realizations and the optimized RC sequence that defines the shuttling trajectory. We then transform each final state into a frame rotating with the spin in the valley groundstate, defined by

\begin{equation}
U(t)=\exp\!\left(-\frac{i}{\hbar}H_{\mathrm{eff}}\,t\right)
\end{equation}
where $H_\text{eff}$ is:

\begin{equation}
H_{\text{eff}} = H_{\text{S}} + \gamma \sigma_z
\end{equation}

where $ \gamma  = -\frac{1}{2} \, \frac{\delta g}{g} \, \lvert B_z \rvert \, \mu_B $ coming from the Zeeman term as described in \cite{davidLongDistanceSpin2024}.

Then the unrotated density matrices of the states $\rho_i $ are obtained in the following way:

\begin{equation}
    \rho_i^{\,\text{(unrotated)}} = U^\dagger(t_i)\, \rho_i \, U(t_i)
\end{equation}

The Bloch components of these states are extracted using expectation values of the Pauli operators as follows: 

\begin{equation}
\mathbf{r}_i
=
\operatorname{Re}\!\left[
\mathrm{Tr}\!\left(\rho_i^{\text{(unrotated)}} \, \boldsymbol{\sigma}\right)
\right],
\qquad
\boldsymbol{\sigma} = (\sigma_x,\sigma_y,\sigma_z).
\end{equation}

From these coordinates, a mean center state is computed:

\begin{equation}
\boldsymbol{r}_{\text{center}}
=
\frac{1}{N_{\text{noise}}}
\sum_{i=1}^{N_{\text{noise}}}
\boldsymbol{r}_i. = (x_\text{center}, y_\text{center}, z_\text{center}).
\end{equation}
This vector is then mapped to a pure quantum state on the Bloch sphere:
\begin{equation}
\theta = \arccos(z_\text{center}), \qquad \phi = \operatorname{atan2}(y_\text{center}, x_\text{center})
\end{equation}
\begin{equation}
\lvert \psi_{\text{center}} \rangle =
\begin{pmatrix}
\cos(\theta/2) \\
e^{i\phi}\sin(\theta/2)
\end{pmatrix}
\end{equation}
where $\theta$ is the polar angle, and $\phi$ is the azimuthal angle. The fidelity between each noise-influenced final state and the center state is computed:
\begin{equation}
F_i = \langle \psi_{\text{center}} \rvert  \rho_i^{\,\text{(unrotated)}} \lvert \psi_{\text{center}} \rangle
\end{equation}
and an average fidelity over the point cloud is then defined as
\begin{equation}
\mathcal{F} = \frac{1}{N_{noise}} \sum_{i=1}^{N_{noise}} F_i
\end{equation}
In this way, the fidelity is a measure of the ensemble coherence, quantifying how tightly the final states cluster around their mean Bloch-sphere direction. 

\section{GA optimizer details}\label{app:optimization}
The computational effort required for the Genetic Algorithm (GA) optimization decreases systematically with increasing shuttling velocity \(v_{\mathrm{avg}}\), as can be seen in Tab.~\ref{tab:complexity}. At low velocities (\(v_{\mathrm{avg}}\le 10\,\mathrm{m\,s^{-1}}\)), convergence to a spin purity of $\bar{P}_s(t_f) > 0.9999$ requires substantially more generations, consistent with a more challenging optimization landscape; increasing \(v_{\mathrm{avg}}\) leads to markedly faster convergence.

This trend cannot be attributed solely to simulation runtime. Since the simulator employs a fixed time grid, slower shuttling implies longer trajectories and thus more transient time steps, so the per-evaluation cost increases approximately linearly as \(v_{\mathrm{avg}}\) decreases. Yet the wall-clock time per GA generation varies only weakly across the velocities studied, indicating that the dominant contribution is the increased number of generations required at low \(v_{\mathrm{avg}}\).

Among the accessible operating points, \(v_{\mathrm{avg}}=20\,\mathrm{m\,s^{-1}}\) is the most efficient case considered here. Across five independent valley disorder realizations, the target objective is reached within \(1\)–\(4\) generations (mean \(2.2\) generations), corresponding to a mean wall-clock optimization time of \(\approx 0.15\)~h. Shuttling velocities above \(20\,\mathrm{m\,s^{-1}}\) are not achievable with the signal generator presented in this work.

\begin{table}[htbp]
\caption{\label{tab:complexity} Mean computational resource requirements for the Genetic Algorithm (GA) optimization across five different valley maps. In this example, a netlist without parasitic effects and without electrical noise was optimized. All runs were performed on a dual-socket AMD EPYC 9354 system (64 physical cores total), running Linux, using the same parallelization settings for all velocities.}
\begin{ruledtabular}
\begin{tabular}{ccc}
\shortstack{\textbf{Velocity} ($v_{avg}$) \\ ($\mathrm{m\,s^{-1}}$)} &
\shortstack{\textbf{Avg. Generations} \\ \textbf{to Converge}} &
\shortstack{\textbf{Avg. Opt. Time} \\ \textbf{(hours)}} \\
\colrule
5  & 16.0  & 0.90 \\
10 & 13.8  & 0.58 \\
15 & 2.8   & 0.18 \\
20 & 2.2   & 0.15 \\
\end{tabular}
\end{ruledtabular}
\end{table}

\section{Shuttling duty cycle estimation}\label{app:duty_cycle}
To contextualize the per-qubit dissipation at the system level, we estimate shuttling activity within a surface-code cycle. Following the Spiderweb processor architecture~\cite{boterSpiderwebArraySparse2022}, the cycle time is
\begin{equation}
t_{\mathrm{SC}} = 22\,t_{\mathrm{shuttle}} + 14\,t_{1q} + 8\,t_{2q} + t_{\mathrm{readout}} \approx 28.2~\mu\mathrm{s},
\end{equation}
using timing assumptions~\cite{kunneSpinBusArchitectureScaling2024} with \(t_{1q}\approx 200~\mathrm{ns}\), \(t_{2q}\approx 80~\mathrm{ns}\), and \(t_{\mathrm{readout}}\approx 10~\mu\mathrm{s}\)~\cite{fuketaCryogenicCMOSCurrent2022}. For shuttling over \(10~\mu\mathrm{m}\) at \(v_{\mathrm{avg}}=20~\mathrm{m\,s^{-1}}\), \(t_{\mathrm{shuttle}}\approx 500~\mathrm{ns}\), yielding the shuttling duty cycle
\begin{equation}
D_{\mathrm{shuttle}} \approx \frac{22\,t_{\mathrm{shuttle}}}{t_{\mathrm{SC}}} \approx 0.39.
\end{equation}

\section{Noise analysis details}\label{app:noise_details}
\begin{figure}[H]
  \centering
  \includegraphics[width=0.9\linewidth]{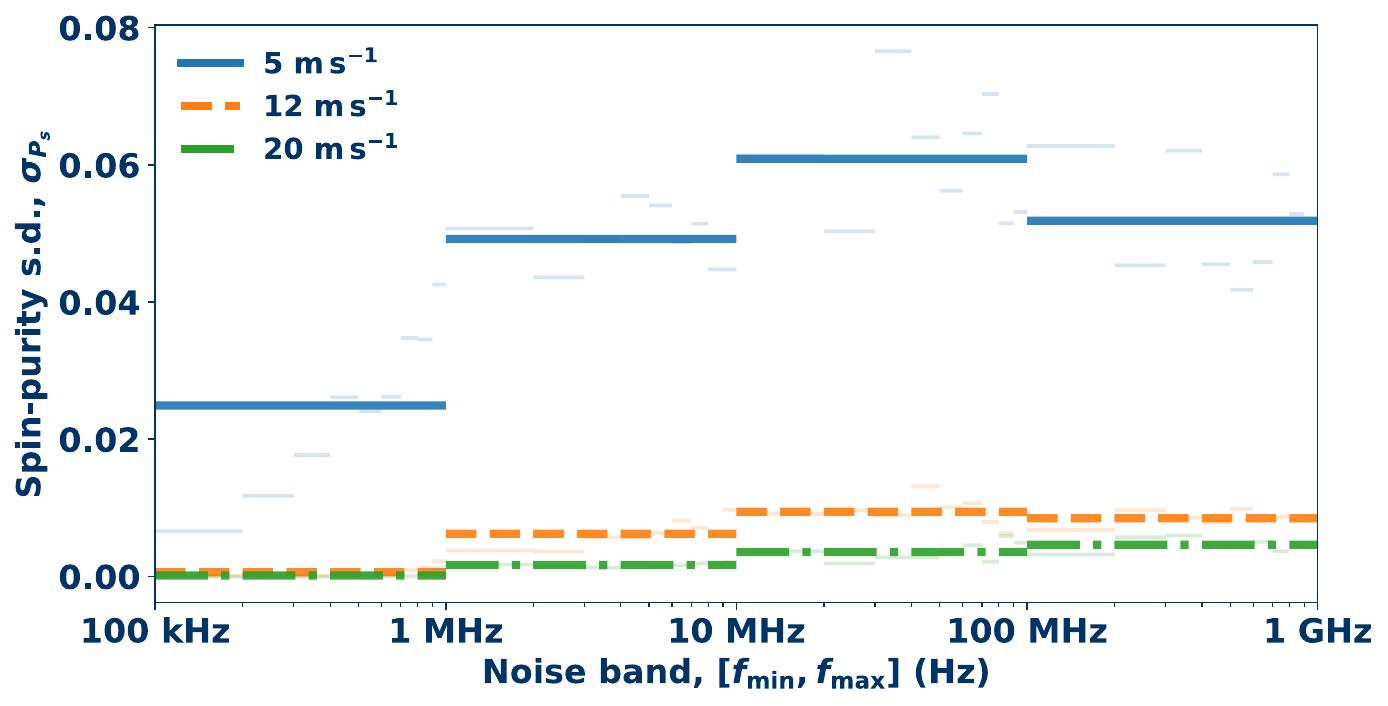}
    \caption{
    Standard deviation of the final spin purity, $\sigma_{P_s}(t_f)$, as a function of the applied noise frequency band for conveyor-mode shuttling over $10~\mu\mathrm{m}$. Results are shown for average shuttling velocities $v_{\mathrm{avg}} = 5$, 12, and $20~\mathrm{m\,s^{-1}}$. For each band, $\sigma_{P_s}(t_f)$ is evaluated over $N_{noise}=50$ independent electronic-noise realizations (shot-to-shot) obtained from \emph{Cadence Spectre} simulations in which device noise is enabled only within the indicated frequency interval. Light segments denote logarithmic sub-band values, while thick bars indicate decade-averaged results.
    }
  \label{fig:purity_over_frequency}
\end{figure}

To identify which noise frequencies contribute most strongly to spin-purity dispersion, we perform a noise-band sweep using transient-noise simulations of the cryogenic signal generator. Fig.~\ref{fig:purity_over_frequency} shows the standard deviation of the purity $\sigma_{P_s}(t_f)$ for a $10~\mu\mathrm{m}$ shuttling operation at $v_{\mathrm{avg}} = 5$, 12, and $20~\mathrm{m\,s^{-1}}$, corresponding to shuttling-signal frequencies of 12.5, 30, and $50~\mathrm{MHz}$, respectively (cf.\ Eq.~\eqref{eq:shuttling_frequency}). For each noise band between $100~\mathrm{kHz}$ and $1~\mathrm{GHz}$, we evaluate $\sigma_{P_s}(t_f)$  over $N_{noise}=50$ independent electronic-noise realizations (shot-to-shot) obtained from \emph{Cadence Spectre} simulations by enabling device noise only within the selected frequency interval. The results show that the dispersion of the final spin purity increases with the activated noise frequency and is substantially reduced at higher shuttling velocities, consistent with increased robustness of faster transport against circuit-noise-induced waveform fluctuations.

The Spectre-based transient-noise simulations include internal device-noise sources as shaped by the circuit transfer function at the chosen operating point. Consequently, Fig.~\ref{fig:purity_over_frequency} should not be interpreted as a device-independent ``universal'' sensitivity curve of the spin--valley dynamics, but rather as a circuit-specific breakdown of how noise contributions from different frequency ranges in the present cryogenic shuttling signal generator translate into run-to-run dispersion.

\section{Electronic simulation details}\label{app:simulation_details}
For the transient circuit simulations we use the following \texttt{Spectre} analysis instruction:

\vspace{0.5pt}
\hspace{-1pt}
\begin{minipage}{\linewidth}
\begin{lstlisting}[style=spectre]
tran <name> stop=<time> method=gear2only strobeperiod=100p maxstep=100p reltol=1e-5 vabstol=1n iabstol=1p noisefmax=10G noisefmin=1 noiseseed=<seed> noise_scale=0.1
\end{lstlisting}
\end{minipage}
Here, \texttt{stop} sets the total simulated time horizon, while \texttt{method=gear2only} enforces a second-order Gear integration scheme that is robust for stiff mixed-signal dynamics. The parameters \texttt{maxstep=100p} and \texttt{strobeperiod=100p} (both 100\,ps) bound and sample the transient solution on a fine, uniform time grid: \texttt{maxstep} prevents the solver from taking larger adaptive steps, and \texttt{strobeperiod} ensures waveform outputs are recorded at deterministic, evenly spaced instants. Tight error controls are imposed via \texttt{reltol=1e-5} together with absolute voltage and current tolerances \texttt{vabstol=1n} and \texttt{iabstol=1p}, limiting numerical jitter and improving reproducibility across runs. Finally, device-noise is enabled over \texttt{noisefmin=1}--\texttt{noisefmax=10G} with a \texttt{noiseseed} that is set individually per noise realization. Using such small time steps is essential when the resulting electrode waveforms are coupled into the quantum-dynamical system: a fixed, sufficiently fine temporal discretization yields deterministic sampling of the control Hamiltonian, avoids solver-dependent variations in the time grid, and thereby suppresses spurious non-determinism (i.e., numerical noise) in the simulated quantum evolution that could otherwise arise purely from adaptive-step interpolation artifacts. The overall noise scaling is set to 0.1. The PDK is evaluated at its characterized ambient temperature of $T\approx 230~\mathrm{K}$, and we use the noise scale parameter to approximate the reduced thermal noise expected under cryogenic operation by scaling the noise amplitude down by a factor of ten.

\FloatBarrier
\bibliographystyle{apsrev4-2}

\end{document}